\documentclass[11pt,letterpaper,draftcls,oneside,peerreview,onecolumn]{IEEEtran}
\usepackage{amsfonts}
\usepackage{amssymb}
\usepackage{amsmath}
\usepackage[dvips]{graphicx}
\usepackage{cite}
\usepackage{hyperref}
\usepackage[left=0.9in,top=1.0in,bottom=1.2in,right=0.9in]{geometry}

\newtheorem{theorem}{Theorem}

\newtheorem{lemma}{Lemma}

\begin{document}

\title{The Diversity Potential of Relay Selection with Practical Channel
Estimation}
\author{Diomidis S. Michalopoulos,~\IEEEmembership{Member,~IEEE,} Nestor D.
Chatzidiamantis,~\IEEEmembership{Student~Member,~IEEE,} Robert Schober,~%
\IEEEmembership{Fellow,~IEEE,} and George K. Karagiannidis,~%
\IEEEmembership{Senior~Member,~IEEE} \thanks{%
This paper was presented in part at the IEEE International
Conference on Communications (ICC), 2011.} \thanks{%
D. S. Michalopoulos and R. Schober are with the Department of Electrical and
Computer Engineering, University of British Columbia, Vancouver, BC V6T1Z4,
Canada, (e-mails: \{dio, rschober\}@ece.ubc.ca)}\thanks{%
N. D. Chatzidiamantis and G. K. Karagiannidis are with the Wireless
Communications Systems Group (WCSG), Department of Electrical and Computer
Engineering, Aristotle University of Thessaloniki, GR-54124 Thessaloniki,
Greece (e-mails: \{nestoras, geokarag\}@auth.gr).}}
\maketitle

\begin{abstract}
We investigate the diversity order of decode-and-forward relay selection
in Nakagami-$m$ fading, in cases where practical channel estimation techniques are applied. In this respect, we introduce a unified model for the imperfect channel estimates, where the effects of noise, time-varying channels, and feedback delays are jointly considered. Based on this
model, the correlation between the actual and the estimated channel
values, $\rho$, is expressed as a function of the signal-to-noise ratio (SNR),
yielding closed-form expressions for the overall outage probability as a function of $\rho$.
The resulting diversity order and power gain reveal a high
dependence of the performance of relay selection on the high SNR behavior of $\rho$, thus shedding light onto the effect of channel estimation on the overall performance. It is shown that when the channel estimates are not
frequently updated in applications involving time-varying channels, or when
the amount of power allocated for channel estimation is not sufficiently high, the
diversity potential of relay selection is severely degraded.

In short, the main contribution of this paper lies in answering the
following question: How fast should $\rho$ tend to one, as the SNR tends to
infinity, so that relay selection does not experience any diversity loss?
\end{abstract}


\newpage

\section{Introduction}

Wireless relaying technology has been recently proposed as a method that
promises significant performance improvement in wireless communications
without any power increase \cite{dohler,uysal}. Among the most common
relaying techniques is so-called relay selection which has been extensively
analyzed in the literature 
\cite{uysal2,beres,zhao,mehta,mehta2}.
In relay selection, the system is able to select a single relay out of the
set of available relays, in order to take advantage of the multiple paths
available and thus achieve spatial diversity. It has been shown that
activating only the relay with the strongest instantaneous end-to-end
channel represents a bandwidth-efficient alternative to all-participate
relaying, since, on the one hand, the same diversity order is achieved, yet
on the other hand, the excessive bandwidth usage that the activation of
multiple relays entails is avoided \cite{zhao}.

Most of the literature dealing with relay selection in fading channels has
assumed that perfect channel state information (CSI) is available at the terminal
where the decision on which relay to activate is made. This assumption, however, may
not be true in practical scenarios where the channel changes rapidly
enough, so that the CSI available at the selecting terminal is outdated. In
addition, if the power allocated to the pilot symbols is not sufficiently
high, the noisy channel estimates may lead to suboptimal relay selection.
The above two cases reveal the vulnerability of relay selection to imperfect
channel estimation, and constitute the main rationale for conducting a
thorough outage and diversity analysis of relay selection in scenarios with
imperfect CSI in this work.

In fact, the case of relay selection under outdated CSI and Rayleigh fading
has been recently studied in \cite{vicario,outCSI,outCSINak,outCSIJ,torabi,soysa,gayan,gayan2}, where
interesting results on the outage probability and diversity order were
derived. Nonetheless, these works consider only a special case, 
since they assume
that the CSI imperfection stems only from delayed feedback. This 
may not always be the case in practice, since channel estimates may also be
impaired by time-varying fading and channel noise. In the very recent works
\cite{seyfi,jafar}, the effect of noisy channel estimates is also included in
the performance analysis. However, these works are based on the assumption that the same
estimates are used for both relay selection and detection, leading to zero
diversity order. Such assumption may not always be true in
practical scenarios where the number of pilots used for relay selection and
that used for symbol detection may not be equal to each other.

In light of the above, the contributions of this paper are summarized as
follows.

\begin{itemize}
\item We conduct an outage analysis of relay selection with
imperfect CSI, which is general enough so as to account for both the effects of noisy
and outdated channel estimates, integrated into a unified model.
In particular, a closed-form expression for the outage probability is
derived, which incorporates all effects that can cause imperfect channel
estimates in practical applications. The considered channel estimation techniques include the cases of estimation in noisy static channels; estimation in noiseless time-varying channels, and estimation in noisy time-varying channel with the aid of finite impulse response (FIR) and infinite impulse response (IIR) channel prediction.

\item The amount of CSI imperfection is reflected by the correlation
coefficient between the actual and the estimated channel values, $\rho$, which is
modeled as a non-decreasing function of the signal-to-noise ratio (SNR). As
a result, the asymptotic outage behavior of relay selection is determined by the
speed of convergence of the correlation coefficient to unity, as the SNR approaches
infinity. 
The diversity order of relay selection with imperfect CSI is thereby derived, shedding light onto the diversity loss caused by imperfect CSI, with implications for the design of channel estimation techniques.

\item In contrast to other relevant works in the literature, where Rayleigh
fading channels were assumed, the versatile scenario of Nakagami-$m$ fading
is studied. It is shown that the resulting diversity order is directly
proportional to the fading shape parameter, $m$.
\end{itemize}

Overall, the general conclusion of this paper is that the level of CSI
imperfection plays an important role in the overall performance of relay
selection, considerably affecting its diversity potential. A detailed
discussion on the diversity order of relay selection for several practical channel estimation techniques is presented in Section \ref{DivPot},
and corresponding numerical examples are given in Section \ref{NR}. These
results are based on the exact outage analysis conducted in Section \ref{PA}
for Rayleigh fading and certain channel estimation techniques, and extended to Nakagami-$m$ fading in Section \ref{PA2}. Prior to the outage
analysis, the unified model that incorporates the effects of noisy and
outdated channel estimates is presented in Section \ref{Imprf}. The system
model of decode-and-forward (DF) relay selection with imperfect CSI is given next, in Section \ref{SM}.

\section{\label{SM}System Model}

Let us consider a cooperative relaying system which consists of a single
source terminal, $S$, $N$ DF relays which are denoted by $R_{i}$, $i=1,...,N$,
and operate in the half-duplex mode \cite{dohler}, and a single destination
terminal, $D$.

\subsubsection*{Channel Model}

Let $h_{AB}$ denote the complex channel between nodes $A$ and $B$, where
$A,B\in \left\{ S,D,R_{i}:\right. $ $\left. i=1,...,N\right\} $. Moreover,
Rayleigh distributed fading in each of the participating links is
assumed, implying that $h_{AB}$ is complex Gaussian random variable (RV).
The versatile scenario of Nakagami-$m$ distribution of fading is considered in
Section \ref{PA2}. In addition, since in this work we focus our attention on
the asymptotic properties of relay selection under imperfect CSI, we assume
independent and identically distributed (i.i.d.) fading in each of the links
involved. Moreover, the fading is considered slow enough such that $h_{AB}$ remains constant during the transmission of one frame.

Let $\gamma _{AB}$ represent the instantaneous SNR of the link between
terminals $A$ and $B$, i.e., $\gamma _{AB}=\left\vert h_{AB}\right\vert
^{2}/N_{0}$, where $N_{0}$ is the additive white Gaussian noise (AWGN)
power. Due to the i.i.d. fading assumption, the average SNR in each of the
links involved is identical, and denoted by $\bar{\gamma}$. Moreover, we use
the notation $f_{X}\left( \cdot \right) $ and $F_{X}\left( \cdot \right) $
to refer to the probability density function (pdf) and the cumulative
distribution function (cdf) of RV $X$, respectively.

\subsubsection*{Relay Selection Process}

Among the available relays, only a single relay is activated in each
transmission session, based on the selection cooperation protocol \cite%
{beres}. In particular, the relay selection procedure is completed in two
phases, as follows. In the first phase, the relays that can successfully
decode the message form the so-called decoding set, denoted by $\mathcal{S}$.

Mathematically speaking, the decoding set $\mathcal{S}$ is defined as $%
\mathcal{S=}\left\{ R_{i}:\gamma _{SR_{i}}>T\right\} $, where $T$ denotes
the outage threshold SNR, defined as the maximum SNR value that allows
decoding; $T$ is related to the target data rate, $r$, through $T=2^{2r}-1$. In
the second phase, the destination collects the estimated CSI of the $R_{i}$-$%
D$ links with $i:R_{i}\in \mathcal{S}$, and activates the relay with the
strongest $R_{i}$-$D$ channel.

A fundamental principle throughout this paper is the fact that the relay
selection is based not on the actual channel values but on their estimates,
which are generally not equal to each other. In this respect, let $\hat{h}%
_{AB}$ denote the estimate of channel $h_{AB}$, so that $\hat{\gamma}%
_{AB}$ represents the estimated value of $\gamma _{AB}$, as seen by the
destination. Hence, denoting the selected relay by $R_{\kappa }$, we have
\begin{equation}
\kappa =\arg \max_{i:R_{i}\in \mathcal{S}}\hat{\gamma}_{R_{i}D}.
\end{equation}%
The CSI imperfection is assumed to affect the relay
selection process, but not the symbol detection at the destination. This is
because the number of pilot symbols used for detection is typically higher
than that used for relay selection, and the channel estimates for detection
can be updated more frequently. A detailed description of the
considered imperfect CSI model follows.

\section{\label{Imprf}Imperfect CSI Model}

The physical causes of the considered CSI degradation are the time-varying
nature of fading channels, as well as finite pilot symbol power. In this
work, both of these causes of imperfect CSI are integrated into a unified
model, as shown below.

The level of CSI imperfection is quantified by the correlation coefficient
between the actual squared channel envelope value, $\left\vert
h_{AB}\right\vert ^{2}$, and its corresponding estimate, $|\hat{h}_{AB}|^{2}$%
, where $A$ and $B$ can be any terminals of the set $\left\{ S,D,R_{i}:\text{
}i=1,...,N\right\} $. This coefficient is defined as (in the sequel, all
channel indices are dropped due to the i.i.d. fading assumption)%
\begin{equation}
\rho =\frac{E\left\langle \left( \left\vert h\right\vert ^{2}-\Omega
_{h}\right) \left( |\hat{h}|^{2}-\Omega _{\hat{h}}\right) \right\rangle }{%
\sigma _{\left\vert h\right\vert ^{2}}\sigma _{|\hat{h}|^{2}}}  \label{rho1}
\end{equation}%
where $\Omega _{h}=E\left\langle \left\vert h\right\vert ^{2}\right\rangle $%
, $\Omega _{\hat{h}}=E\left\langle |\hat{h}|^{2}\right\rangle $ with $%
E\left\langle \cdot \right\rangle $ denoting expectation, and $\sigma _{X}$
denotes the standard deviation of RV $X$.

Note that $\rho $ reflects the effect of imperfect CSI on the SNR of the
selected relay and hence on the overall performance of relay
selection with imperfect CSI. For this reason, the subsequent analysis focuses on expressing
the performance degradation as a function of $\rho $, so that all physical
phenomena that cause CSI degradation are, in fact, incorporated into $\rho $.

\subsection{Versatile Imperfect CSI Case}

In accordance with the intuition that channel estimation is performed in
noisy environments, let us consider the versatile scenario where $\rho $ is a
function of the SNR. That is,%
\begin{equation}
1-\rho =g\left( \bar{\gamma}\right)  \label{g1}
\end{equation}%
where $g\left( \cdot \right) $ is generally a non-increasing function of its
argument, with $0\leq g\left( \bar{\gamma}\right) \leq 1$. In order to
obtain insight into the asymptotic dependence of the CSI error on the SNR,
we expand $g\left( \bar{\gamma}\right) $ into a Puiseux series \cite{cherlin}%
, so that for high SNR we have%
\begin{equation}
g\left( \bar{\gamma}\right) =b\bar{\gamma}^{-a}+o\left( \bar{\gamma}%
^{-a}\right)  \label{grd}
\end{equation}%
where $a$ and $b$ are positive constants with $0<b\leq 1$, and $o\left( \bar{\gamma%
}^{-a}\right) $ is defined such that $\lim_{\bar{\gamma}\rightarrow \infty
}o\left( \gamma ^{-a}\right) /\left( \bar{\gamma}^{-a}\right) =0$.

It is emphasized that the versatile imperfect CSI model considered in (\ref%
{g1}) and (\ref{grd}) is general enough to accommodate the cases of
imperfect CSI due to noise impairment and the time-varying nature of the
underlying channels. Next, we study the scenarios of CSI imperfection in
static and time-varying Rayleigh fading channels, separately.

\subsection{\label{Static}Static Channels, Noisy CSI}

Let us assume the case of static channels, where channel estimation is
implemented via averaging over $L$ noisy pilot symbols. As a result, $%
g\left( \bar{\gamma}\right) $ in (\ref{g1}) is a decreasing function of $%
\bar{\gamma}$, for which $\lim_{\bar{\gamma}\rightarrow \infty }g\left( \bar{%
\gamma}\right) =0$ holds. Moreover, let us consider the scenario where the
power allocated to pilot symbols, $\mathcal{E}_{p}$, is not necessarily
equal to the power allocated to data transmission, $\mathcal{E}_{d}$. We
allow the ratio of $\mathcal{E}_{p}$ over $\mathcal{E}_{d}$ to be
SNR-dependent, so that%
\begin{equation}
\mathcal{E}_{p}=\beta \bar{\gamma}^{\alpha }\mathcal{E}_{d}  \label{Ep}
\end{equation}%
where $\beta $ is a positive constant and $\alpha $ is a constant, the sign
of which determines whether $\mathcal{E}_{p}$ increases or decreases with
SNR. The estimated channel values are expressed as $\hat{h}=h+n_{p}
$, where $h$ and $n_{p}$ denote the true channel component and
the remaining noise component, respectively. Given that the channel
estimates are derived by averaging over $L$ pilot symbols, the noise
variance of the estimation process equals $\sigma _{n_{p}}^{2}=N_{0}/\left(
\mathcal{E}_{p}L\right) .$

\begin{lemma}
\label{rhoT}The correlation coefficient, $\rho $, between $\left\vert
h\right\vert ^{2}$ and $|\hat{h}|^{2}$, is given by%
\begin{equation}
\rho =\frac{\Omega _{h}}{\Omega _{\hat{h}}}=\frac{\Omega _{h}}{\Omega
_{h}+\sigma _{n_{p}}^{2}}=\frac{L\beta \bar{\gamma}^{\alpha +1}}{L\beta \bar{%
\gamma}^{\alpha +1}+1}.  \label{Theo}
\end{equation}
\end{lemma}

\begin{proof}
Since $\hat{h}$ equals a linear combination of complex Gaussian RVs, $|\hat{h%
}|^{2}$ is exponentially distributed. Hence, it follows from the theory of
the moments of exponential RVs that $E\left\langle \left\vert h\right\vert
^{4}\right\rangle =2\Omega _{h}^{2}$; $E\left\langle |\hat{h}%
|^{4}\right\rangle =2\Omega _{\hat{h}}^{2}$. Using this result, the proof
follows from (\ref{rho1}) after algebraic manipulations, in conjunction with
(\ref{Ep}) and the fact that $\bar{\gamma}=\mathcal{E}_{d}\Omega _{h}/N_{0}$.
\end{proof}

Expanding (\ref{Theo}) in a Taylor series for $\bar{\gamma}\rightarrow
\infty $ and using (\ref{g1}), we obtain for $\alpha >-1$\footnote{%
The case of $\alpha <-1$ yields $\rho =0$ for $\bar{\gamma}\rightarrow
\infty $, and is out of the scope of this paper.}%
\begin{equation}
g\left( \bar{\gamma}\right) =1-\rho =\frac{1}{\beta L}\bar{\gamma}%
^{-\left( \alpha +1\right) }+o\left( \bar{\gamma}^{-\left( \alpha +1\right)
}\right) .  \label{rhoA2}
\end{equation}%
Therefore, using (\ref{grd}), from (\ref{rhoA2}) we have $a=\alpha +1$; $%
b=1/\left(\beta L\right)$.

\subsection{Time-Varying Channels}

Next, the case of time-varying Rayleigh fading is studied, where the maximum Doppler
frequency on each of the participating links is assumed identical, and
denoted by $f_{d}$. Moreover, the autocorrelation function of the complex
channel $h$ is denoted by $\rho _{h}\left( T_{d}\right) $; based on the
Jakes' model \cite{Jakes}, $\rho _{h}\left( T_{d}\right) $ is given as $\rho
_{h}\left( T_{d}\right) =\Omega _{h} J_{0}(2\pi f_{d}T_{d})$, where $%
J_{0}\left( \cdot \right) $ denotes the zeroth order Bessel function of the
first kind \cite[Eq. (8.411)]{Gradshteyn}.

\subsubsection{FIR Channel Prediction}

In time-varying environments, the channel estimation can be improved by
utilizing the CSI available from previous time instances, so that the
channel estimates are derived through a \textit{channel prediction }process
\cite{makhoul}. Let us consider an FIR channel prediction filter of length $%
L $, and denote the time interval between consecutive CSI acquisitions by $%
T_{d}$. In such case, following the analysis in \cite{makhoul}, the
predictor coefficients can be optimized so as to yield the minimum squared
error between the actual and the predicted channel values, $\sigma _{e}^{2}$%
, resulting in%
\begin{equation}
\sigma _{e}^{2}=\Omega _{h}-\text{\b{u}}_{h}^{H}\text{\b{R}}^{-1}\text{\b{u}}%
_{h}.  \label{pvec}
\end{equation}%
In (\ref{pvec}), \b{u}$_{h}$ denotes the $L$-dimensional autocorrelation
vector, i.e., \b{u}$_{h}=\left[ \rho _{h}\left( -T_{d}\right) ,...,\rho
_{h}\left( -LT_{d}\right) \right] ^{T}$, \b{R} denotes an $L\times L$
symmetric Toeplitz matrix, the first row of which is given by $\left[ \rho
_{h}\left( 0\right) +N_{0}/\mathcal{E}_{p},\rho _{h}\left( -T_{d}\right)
,...,\right.$
\\$\left.\rho _{h}\left( -LT_{d}+T_{d}\right) \right] $, and $\left( \cdot
\right) ^{H}$ denotes the Hermitian operator. Hence, $\rho $ is derived by
combining (\ref{rho1}) and (\ref{pvec}), as%
\begin{equation}
\rho =\text{\b{u}}_{h}^{H}\text{\b{R}}^{-1}\text{\b{u}}_{h}/\Omega _{h}.
\label{rhoFIR}
\end{equation}%
It follows from (\ref{g1}) that the CSI error can be expressed as a
function of the SNR as $g\left( \bar{\gamma}\right) =1-\rho =1-$\b{u}$%
_{h}^{H}$\b{R}$^{-1}$\b{u}$_{h}/\Omega _{h}$. Interestingly, it is noted
that in the high-SNR regime and for $f_{d}>0$, $g\left( \bar{\gamma}\right) $
converges to a finite non-zero constant, i.e.,%
\begin{equation}
\lim_{\bar{\gamma}\rightarrow \infty }g\left( \bar{\gamma}\right) =1-\text{%
\b{u}}_{h}^{H}\left. \text{\b{R}}^{-1}\right\vert _{N_{0}=0}\text{\b{u}}%
_{h}/\Omega _{h}=b>0
\end{equation}%
implying that the CSI error is independent of the SNR. Hence, considering (%
\ref{grd}), it follows that for the case where the channel estimates are
obtained through FIR channel prediction, $a=0$ holds.

\paragraph*{\label{outdatedCSI}Ideal but Outdated CSI}

This special case of channel estimation was considered in \cite%
{vicario,outCSI,soysa,outCSINak,outCSIJ,torabi,gayan}, and in fact
corresponds to noiseless FIR channel prediction with a one-tap predictor $%
\left( L=1\right) $, and is dubbed as \textquotedblleft outdated
CSI\textquotedblright\ here. It implies that the CSI based on which the
\textquotedblleft best\textquotedblright\ relay is selected is noise-free,
yet the selection of the \textquotedblleft best\textquotedblright\ relay is
not based on the current time instant but on a previous one, because of,
e.g., a feedback delay. Based on (\ref{rhoFIR}), it can be shown that the
correlation coefficient, $\rho $, for the outdated CSI case equals $\rho
=\rho _{h}^{2}\left(-T_{d}\right) /\Omega _{h}^{2}$, a result which is in
accordance with \cite{vicario}, \cite{outCSI}. Moreover, $g\left( \bar{\gamma%
}\right) $ is a constant function in this case, and thus (\ref{grd}) yields $%
a=0$; $b=1-\rho _{h}^{2} \left(-T_{d}\right) /\Omega _{h}^{2}$.

\subsubsection{IIR Channel Prediction}

Let us now extend the channel prediction case to the scenario where the
number of pilot symbols participating in the prediction process are
infinitely large. As shown in Appendix \ref{ApIIR}, this scenario leads to a
correlation coefficient of%
\begin{equation}
\rho =1-\exp \left( T_{d}\int_{-f_{d}}^{f_{d}}\ln \left[ \mathcal{S}%
_{hh}\left( e^{j2\pi fT_{d}}\right) +\left( \beta \bar{\gamma}^{\alpha
+1}\right) ^{-1}\right] df\right) \left( \beta \bar{\gamma}^{\alpha
+1}\right) ^{-\left( 1-2f_{d}T_{d}\right) }+\left( \beta \bar{\gamma}%
^{\alpha +1}\right) ^{-1}  \label{rhosqIIR}
\end{equation}%
where $\mathcal{S}_{hh}\left( \cdot \right) $ represents the Fourier
transform of $\rho _{h}\left( \cdot \right) $. Hence, combining (\ref{g1})
and (\ref{rhosqIIR}), we obtain for high SNR%
\begin{equation}
g\left( \bar{\gamma}\right) =\underset{b}{\underbrace{\exp \left(
T_{d}\int_{-f_{d}}^{f_{d}}\ln \left[ \mathcal{S}_{hh}\left( e^{j2\pi
fT_{d}}\right) \right] df\right) \beta ^{-\left( 1-2f_{d}T_{d}\right) }}}%
\bar{\gamma}^{-\underset{a}{\underbrace{\left( \alpha +1\right) \left(
1-2f_{d}T_{d}\right) }}}.  \label{limIIR2}
\end{equation}%
Consequently, it is concluded that the parameters $a$ and $b$ of the
asymptotic dependence of the CSI error on the SNR are given by $a=\left(
\alpha +1\right) \left( 1-2f_{d}T_{d}\right) $ and $b=\exp \left(
T_{d}\int_{-f_{d}}^{f_{d}}\ln \left[ \mathcal{S}_{hh}\left( e^{j2\pi
fT_{d}}\right) \right] df\right) \beta ^{-\left( 1-2f_{d}T_{d}\right) }$.

The reader is referred to Table \ref{Table} for an overview of how the
parameters $a$ and $b$ are derived for the practical channel estimation
scenarios considered in this paper. An asymptotic performance analysis of
suboptimal relay selection follows.

\section{\label{PA}Outage Analysis of Relay Selection with Imperfect CSI in Rayleigh
Fading}


The outage probability is defined as the probability that the overall SNR
lies below a given threshold, denoted here by $T$, i.e., $P_{out}=\Pr
\left\{ \gamma _{\kappa }<T\right\} $, where $\kappa $ denotes the index of
the selected relay and $\gamma _{\kappa }$ is the corresponding end-to-end
SNR. Observing that for all channel estimation scenarios considered in
Section \ref{Imprf}, $\hat{h}$ is obtained as a linear combination of complex
Gaussian RVs, it follows that $\hat{h}$ is also a complex Gaussian RV.
Hence, $\hat{\gamma}$ is exponentially distributed. Consequently, the
conditional pdf of the actual SNR, $\gamma $, conditioned on its estimate, $%
\hat{\gamma}$, is obtained from \cite[Eq. (2.11)]{dowton} as%
\begin{equation}
f_{\gamma \left\vert \hat{\gamma}\right. }\left( x\left\vert y\right.\right) =\frac{\exp
\left( -\frac{x}{\bar{\gamma}\left( 1-\rho \right) }-\frac{y\rho }{\overline{%
\hat{\gamma}}\left( 1-\rho \right) }\right) }{\bar{\gamma}\left( 1-\rho
\right) }I_{0}\left( 2\frac{\sqrt{\rho xy}}{\left( 1-\rho \right) \sqrt{\bar{%
\gamma}\overline{\hat{\gamma}}}}\right)  \label{fcond}
\end{equation}%
where $\overline{\hat{\gamma}}$ denotes the average estimated SNR and $%
I_{0}\left( \cdot \right) $ denotes the zeroth order modified Bessel
function of the first kind \cite[Eq. (8.447.1)]{Gradshteyn}. It is
emphasized that since the parameters $\bar{\gamma}$ and $\overline{\hat{%
\gamma}}$ are not necessarily equal to each other, which is in contrast to the outdated CSI case treated in \cite
{vicario,outCSI}, the diversity
investigation of relay selection with imperfect CSI under the general imperfect CSI
assumption requires that we conduct a new outage analysis for our scheme.
This outage analysis is similar to that in \cite{vicario}, yet the corresponding
expression for the $f_{\gamma \left\vert \hat{\gamma}\right. }\left( x\left\vert y\right.\right)$ used in \cite{vicario} is substituted by (\ref{fcond}) .

In particular, based upon the mode of operation of the selection cooperation
\cite{beres} the outage probability is expressed as%
\begin{equation}
P_{out}=\Pr \left\{ \mathcal{S}=\emptyset \right\} +\sum_{l=1}^{N}F_{\gamma
_{R_{\kappa }D}}\left( T\left\vert ~\left\vert \mathcal{S}\right\vert
=l\right. \right) \Pr \left\{ \left\vert \mathcal{S}\right\vert =l\right\}
\label{Pout}
\end{equation}%
where $\left\vert \mathcal{S}\right\vert $ denotes the cardinality of $%
\mathcal{S}$ and $\emptyset $ is the empty set. Because of the i.i.d.
assumption for the fading in the $S$-$R_{i}$ links, the second term within the
sum in (\ref{Pout}) is given by \cite[Eq. (7)]{vicario}
\begin{equation}
\Pr \left\{ \left\vert \mathcal{S}\right\vert =l\right\} =\binom{N}{l}\left[
1-\exp \left( -\frac{T}{\bar{\gamma}}\right) \right] ^{N-l}\exp \left( -%
\frac{lT}{\bar{\gamma}}\right) .  \label{Card1}
\end{equation}%
Furthermore, by defining $A_{i}$ as the event that the $i$th relay out of
$l$ relays is selected, i.e., $A_{i}:i=\kappa $, $F_{\gamma _{R_{\kappa
}D}}\left( T\left\vert ~\left\vert \mathcal{S}\right\vert =l\right. \right) $
is expressed as%
\begin{eqnarray}
F_{\gamma _{R_{\kappa }D}}\left( T\left\vert ~\left\vert \mathcal{S}%
\right\vert =l\right. \right) &=&\sum_{i=1}^{l}F_{\gamma _{R_{\kappa
}D}}\left( T\left\vert ~\left\vert \mathcal{S}\right\vert =l,~A_{i}\right.
\right) \Pr \left\{ A_{i}\right\}  \notag \\
&=&\int_{0}^{T}\int_{0}^{\infty }f_{\gamma _{R_{i}D}\left\vert \hat{\gamma}%
_{R_{i}D}\right. }\left( x_{1}\left\vert x_{2}\right. \right) f_{\hat{\gamma}%
_{R_{i}D}\left\vert A_{i}\right. }\left( x_{2}\left\vert A_{i}\right.
\right) dx_{2}dx_{1}  \label{Cond1}
\end{eqnarray}
where we used the fact that $\Pr \left\{ A_{i}\right\} =1/l$, because of
symmetry. The conditional density of $\hat{\gamma}_{R_{i}D}$ conditioned on
$A_{i}$ is derived as%
\begin{equation}
f_{\hat{\gamma}_{R_{i}D}\left\vert A_{i}\right. }\left( x_{2}\left\vert
A_{i}\right. \right) =\frac{
f_{\hat{\gamma}_{R_{i}D}\left\vert A_{i}\right. }\left( x_{2}\cap
A_{i} \right)}{\Pr \left\{ A_{i}\right\} }=\frac{f_{\hat{\gamma}%
_{R_{i}D}}\left( x_{2}\right) }{\Pr \left\{ A_{i}\right\} }\prod_{\substack{ %
j=1  \\ j\not=i}}^{l}F_{\hat{\gamma}_{R_{j}D}}\left( x_{2}\right) =lf_{\hat{%
\gamma}_{R_{i}D}}\left( x_{2}\right) F_{\hat{\gamma}_{R_{i}D}}^{l-1}\left(
x_{2}\right) .  \label{Fs3}
\end{equation}%
Consequently, substituting (\ref{fcond}) and (\ref{Fs3}) in (\ref{Cond1}) we
obtain an expression for $F_{\gamma _{R_{\kappa }D}}\left( T\left\vert
~\left\vert \mathcal{S}\right\vert =l\right. \right) $ which coincides with
\cite[Eq. (8)]{vicario}, where the case of outdated CSI was considered. This
leads to an interesting observation which is summarized below.

Under i.i.d. Rayleigh fading and assuming correlation coefficient $\rho $
between the actual and the estimated SNR in each intermediate link\textit{,
the outage probability of relay selection with imperfect CSI, }$P_{out}$, \textit{%
expressed as a function of }$\rho $,\textit{\ is given by the same formula,
irrespective of the channel estimation technique used}. Equivalently, $%
P_{out}$ is independent of $\overline{\hat{\gamma}}$, a fact which can be
explained by noting that only the relative values of $\hat{\gamma}_{i}$
are relevant for relay selection, not their absolute values. Hence, scaling all
the estimated SNRs by the same factor does not affect the relay selection
process. Therefore, the outage probability of suboptimal DF relay selection
for any channel estimation technique is as shown in \cite[Eq. (2)]{vicario}.
It is emphasized, however, that \textit{different channel estimation
techniques lead to different dependences of }$\rho $\textit{\ on the SNR,
resulting ultimately in different diversity behaviors.} The diversity order
of relay selection with imperfect CSI will be studied in detail in Section \ref{DIV}.

\section{\label{PA2}Outage Analysis in Nakagami-$m$ Fading}

Let us now consider the case where the fading in all channels follows the
Nakagami-$m$ distribution \cite{nakagami}. In this case, since the
distribution of $\hat{h}$ is unknown for the unified imperfect CSI model, we
confine ourselves to investigating the performance of relay selection with imperfect CSI for the three special cases presented below.

\subsection{Time-Varying Channels: Outdated CSI}

Recall from Section \ref{outdatedCSI} that this case corresponds to
noiseless FIR channel prediction with one tap. Consequently, $\hat{h}$
represents a delayed version of $h$, and $\hat{h}$ follows the same
distribution as $h$, so that the joint pdf of $\gamma $ and $\hat{\gamma}$
is obtained from the bivariate Gamma distribution \cite{Bivariate_Nakagami},
which is simplified using \cite[Eq. (9.210/1)]{Gradshteyn} to%
\begin{equation}
f_{\hat{\gamma},\gamma }\left( x_{1},x_{2}\right) =\sum_{j=0}^{\infty }\frac{%
\left( m\right) _{j}m^{2m+2j}\rho ^{j}}{j!\left( 1-\rho \right) ^{m+2j}}%
\prod_{i=1}^{2}\frac{x_{i}^{m+j-1}\exp \left\{ -\frac{mx_{i}}{\left( 1-\rho
\right) \bar{\gamma}}\right\} }{\Gamma \left( m+j\right) \bar{\gamma}^{m+j}}
\label{Bivar2}
\end{equation}%
where $\left( x\right) _{y}$ denotes the Pochhamer symbol defined in \cite[%
pp. xliii]{Gradshteyn}.

Following the same steps as in (\ref{Pout})-(\ref{Fs3}), and using the fact
that, the pdf and cdf of the SNR for the Nakagami-$m$ fading model are given
by $f_{\gamma }\left( \gamma \right) =m^{m}\gamma ^{m-1}/\left[ \bar{\gamma}%
^{m}\Gamma (m)\right] \exp \left( -m\gamma /\bar{\gamma}\right) $ and $%
F_{\gamma }\left( \gamma \right) =1-\Gamma \left( m,\frac{m}{\bar{\gamma}}%
y\right) /\Gamma \left( m\right) $, respectively, in conjunction with the
binomial expansion \cite[Eq. (1.110)]{Gradshteyn}, we obtain the equivalent
expression for (\ref{Cond1}), pertaining to Nakagami-$m$ fading, as%
\begin{align}
F_{\gamma _{R_{\kappa }D}}\left( T\left\vert ~\left\vert \mathcal{S}%
\right\vert =l\right. \right) & =l\int_{0}^{T}\sum_{j=0}^{\infty
}\frac{\rho ^{j}m^{2m+2j}}{j!(1-\rho )^{m+2j}}\frac{x_{1}^{m+j-1}}{\Gamma
\left( m\right) \Gamma \left( m+j\right) \bar{\gamma}^{2\left( m+j\right) }}%
\exp \left( -\frac{mx_{1}}{(1-\rho )\bar{\gamma}}\right)  \notag \\
& \times \sum_{i=0}^{l-1}\binom{l-1}{i}\left( \frac{-1}{\Gamma \left(
m\right) }\right) ^{i}\left( \int_{0}^{\infty }x_{2}^{m+j-1}\exp \left( -%
\frac{mx_{2}}{\bar{\gamma}\left( 1-\rho \right) }\right) \Gamma \left( m,%
\frac{m}{\bar{\gamma}}x_{2}\right) ^{i}dx_{2}\right) dx_{1}.  \label{FS7}
\end{align}%
It is observed from (\ref{FS7}) that in order to derive a closed-form
expression for $F_{\gamma _{R_{\kappa}D}}\left( T\left\vert ~\left\vert \mathcal{S}%
\right\vert =l\right. \right) $, the following
integral needs to be solved%
\begin{equation}
I\left( \mu ,\alpha ,m,\beta ,j\right) =\int_{0}^{\infty }x^{\mu }\exp
\left( -\alpha x\right) \Gamma ^{j}\left( m,\beta x\right) dx.  \label{I1}
\end{equation}%
For the case of $m\in \mathbb{Z},$ the integral in (\ref{I1}) is evaluated
as illustrated in Appendix \ref{ApI}. Hence, $F_{\gamma _{R_{\kappa}D}}\left( T\left\vert ~\left\vert \mathcal{S}%
\right\vert =l\right. \right) $ can be derived, using (\ref{I2}), \cite[Eq. (3.351/1)]%
{Gradshteyn}, and \cite[Eq. (8.352/6)]{Gradshteyn} as
\begin{align}
F_{\gamma _{R_{\kappa }D}}\left( T\left\vert ~\left\vert \mathcal{S}%
\right\vert =l\right. \right) & =l\sum_{j=0}^{\infty }\frac{\rho
^{j}\left(
\Gamma \left( m+j\right) -\Gamma \left( m+j,\frac{mT}{(1-\rho )\bar{\gamma}}%
\right) \right)}{j!(1-\rho )^{j}\Gamma \left( m\right) \Gamma \left( m+j\right) } 
\sum_{i=0}^{l-1}\binom{l-1}{i}\left( -1\right) ^{i}\psi \left(
m+j,\rho ,i\right)  \label{FS9}
\end{align}%
where we have set%
\begin{equation}
\psi \left( x,\rho ,i\right) =\frac{\Gamma \left( i+1\right) }{\left( \frac{1%
}{\left( 1-\rho \right) }+i\right) ^{x}}\sum_{\substack{ \xi _{0},\xi
_{1},...,\xi _{m-1}=0  \\ \xi _{0}+\xi _{1}+...+\xi _{m-1}=i}}^{i}\left(
\prod\limits_{k=0}^{m-1}\frac{\left( \frac{1}{k!\left( \frac{1}{\left(
1-\rho \right) }+j\right) ^{i}}\right) ^{\xi _{k}}}{\Gamma \left( \xi
_{k}+1\right) }\right) \left( x-1+\sum\limits_{q=0}^{m-1}q\xi _{q}\right) !.
\label{coef}
\end{equation}%
Consequently, a closed-form expression for the outage probability is
obtained by combining (\ref{Pout}), (\ref{Card1}), and (\ref{FS9}), yielding%
\begin{align}
P_{out}& =\left( 1-\frac{\Gamma \left( m,\frac{m}{\bar{\gamma}}T\right) }{%
\Gamma \left( m\right) }\right) ^{N}+\sum_{j=0}^{\infty
}\sum_{l=1}^{N}\sum_{i=0}^{l-1}l\binom{N}{l}\left( 1-\frac{\Gamma \left( m,%
\frac{m}{\bar{\gamma}}T\right) }{\Gamma \left( m\right) }\right)
^{N-l}\left( \frac{\Gamma \left( m,\frac{m}{\bar{\gamma}}T\right) }{\Gamma
\left( m\right) }\right) ^{l}  \notag \\
& \times \frac{\rho ^{j}\left( -1\right) ^{i}}{j!(1-\rho )^{j}\Gamma \left(
m\right) \Gamma \left( m+j\right) }\left( \Gamma \left( m+j\right) -\Gamma
\left( m+j,\frac{mT}{(1-\rho )\bar{\gamma}}\right) \right) \binom{l-1}{i}%
\psi \left( m+j,\rho ,i\right) .  \label{Final}
\end{align}%
It is noted that, for practical SNR values, i.e., $\bar{\gamma}\leq 30$dB,
the infinite series in (\ref{Final}) converges after a finite number of
terms, not greater than 100.

\subsection{Time-Varying Channels: FIR Channel Prediction with Large $L$ and IIR Channel Prediction}

In this case, the channel estimate $\hat{h}$ is obtained as the weighted sum
of a large number of observations in time-varying fading scenarios ($f_d>0$) \cite{makhoul}. Hence, it
follows from the central limit theorem that $\hat{h}$ is a complex Gaussian
RV, so that $|\hat{h}|^{2}$ is exponentially distributed with average value
denoted by $\overline{\hat{\gamma}}$.

The joint pdf of $\gamma $ and $\hat{\gamma}$ is obtained from \cite[Eq. (10)%
]{Bivariate_Nakagami} by setting $m_{1}=m$ and $m_{2}=1$, leading to%
\begin{eqnarray}
&&f_{\hat{\gamma},\gamma }\left( x_{1},x_{2}\right) =\left( 1-\rho \right)
^{m}\sum_{j=0}^{\infty }\frac{\Gamma \left( j+1\right) \rho ^{j}}{j!}\left(
\frac{1}{\overline{\hat{\gamma}}\left( 1-\rho \right) }\right) ^{j+1}\left(
\frac{m}{\bar{\gamma}\left( 1-\rho \right) }\right) ^{m+j}  \notag \\
&&\times \frac{x_{2}^{j}x_{1}^{j+m-1}}{\Gamma \left( j+1\right) \Gamma
\left( j+m\right) }\exp \left( -\frac{1}{1-\rho }\left( \frac{x_{2}}{%
\overline{\hat{\gamma}}}+\frac{mx_{1}}{\bar{\gamma}}\right) \right)
~_{1}F_{1}\left( m-1;m+j;\frac{\rho mx_{1}}{\bar{\gamma}\left( 1-\rho
\right) }\right)  \label{Bivar3}
\end{eqnarray}%
where $_{1}F_{1}\left( \cdot ;\cdot ;\cdot \right) $ is the confluent
Hypergeometric function defined in \cite[Eq. (9.210/1)]{Gradshteyn}.
Following the same procedure as in (\ref{Pout})-(\ref{Fs3}), we obtain the conditional
cdf of $\gamma _{R_{\kappa }D}$ as%
\begin{eqnarray}
F_{\gamma _{R_{\kappa }D}}\left( T\left\vert ~\left\vert \mathcal{S}%
\right\vert =l\right. \right) &=&l\sum_{j=0}^{\infty
}\sum_{i=0}^{l-1}\frac{\binom{l-1}{i}\left( -1\right) ^{i}\left( \frac{m}{%
\bar{\gamma}}\right) ^{m+j}\left( \frac{\rho }{1-\rho }\right) ^{j}}{\left[
1+i\left( 1-\rho \right) \right] ^{j+1}\Gamma \left( m+j\right) }  \notag \\
&&\times \int_{0}^{T}x_{1}^{m+j-1}\exp \left( -\frac{mx_{1}}{\left( 1-\rho
\right) \bar{\gamma}}\right) ~_{1}F_{1}\left( m-1;m+j;\frac{\rho mx_{1}}{%
\bar{\gamma}\left( 1-\rho \right) }\right) dx_{1}  \label{hyper}
\end{eqnarray}%
where $\rho $ is given in (\ref{rhoFIR}) and (\ref{rhosqIIR}). It is
observed that, for the same reasons addressed in Section \ref{PA}, $%
F_{\gamma _{R_{\kappa }D}}\left( T\left\vert ~\left\vert \mathcal{S}%
\right\vert =l\right. \right) $ is independent of $\overline{%
\hat{\gamma}}$.

In case of $m=1$, using \cite[Eq. (3.351/1)]{Gradshteyn}, \cite[Eq. (8.352/2)%
]{Gradshteyn} and the fact that $_{1}F_{1}\left( 0;b;z\right) =1$, (\ref%
{hyper}) reduces after some algebraic manipulations to
\begin{equation}
F_{\gamma _{R_{\kappa }D}}\left( T\left\vert ~\left\vert \mathcal{S}%
\right\vert =l\right. \right) =l\sum_{i=0}^{l-1}\frac{\binom{l-1%
}{i}\left( -1\right) ^{i}\left( 1-\rho \right) }{\left[ 1+i\left( 1-\rho
\right) \right] }\sum_{j=0}^{\infty }\left( \frac{\rho }{1+i\left( 1-\rho
\right) }\right) ^{j}\left[ 1-\sum_{t=0}^{\infty }\frac{\left( \frac{T}{\left( 1-\rho \right) \bar{\gamma%
}}\frac{\rho }{1+i\left( 1-\rho \right) }\right) ^{t}}{t!~e^{\frac{T}{\left( 1-\rho \right) \bar{\gamma}}}}\right] .
\label{m1_1}
\end{equation}%
As a cross check, it follows from \cite[Eq. (0.231)]{Gradshteyn} and the
infinite series representation of the exponential function \cite[Eq.
(1.211/1)]{Gradshteyn}, that (\ref{m1_1}) is equivalent to \cite[Eq. (8)]%
{vicario}. In case of $m>1$, (\ref{hyper}) in conjunction with \cite[Eq.
(3.351/1)]{Gradshteyn} yields
\begin{eqnarray}
F_{\gamma _{R_{\kappa }D}}\left( T\left\vert ~\left\vert \mathcal{S}%
\right\vert =l\right. \right)  &=&l\sum_{j=0}^{\infty
}\sum_{i=0}^{l-1}\sum\limits_{k=0}^{\infty }\frac{\binom{l-1}{i}\left(
-1\right) ^{i}\left( 1-\rho \right) ^{m}\rho ^{i+j}}{\left[ 1+i\left( 1-\rho
\right) \right] ^{j+1}\Gamma \left( m-1\right) }\frac{\Gamma \left(
m-1+k\right) }{\Gamma \left( m+j+k\right) k!}  \notag \\
&&\times \left( \Gamma \left( m+j+k\right) -\Gamma \left( m+j+k,\frac{mT}{%
\left( 1-\rho \right) \bar{\gamma}}\right) \right) .  \label{FIRm}
\end{eqnarray}%
The overall outage probability follows then from (\ref{FIRm}) (or (\ref{m1_1}%
), if $m=1$) and (\ref{Pout}).

\subsection{Static Channels: Noisy CSI in the high SNR Regime}

Under the high SNR assumption, it is valid to assume that $\hat{h}=h+n_{p}
$ is also Nakagami-$m$ distributed, with $\overline{%
\hat{\gamma}}=\bar{\gamma}$. Consequently, the outage probability for this
scenario is given by (\ref{Final}), where $\rho $ is given in
the ensuing Lemma.

\begin{lemma}
The correlation coefficient between $|h|$ and $|\hat{h}|$ for Nakagami-$%
m $ fading is given by%
\begin{equation}
\rho =\frac{\sqrt{\frac{1}{m}\Omega _{h}^{2}}}{\sqrt{\frac{1}{m}\Omega
_{h}^{2}+\sigma _{n_{p}}^{4}+2\Omega _{h}\sigma _{n_{p}}^{2}}}=\frac{L}{%
\sqrt{L^{2}+\frac{2mL}{\beta \bar{\gamma}^{\alpha +1}}+\frac{m}{\beta ^{2}%
\bar{\gamma}^{2\left( \alpha +1\right) }}}}.  \label{rhom}
\end{equation}
\end{lemma}

\begin{proof}
The proof is similar to that for \textit{Lemma} \ref{rhoT} by using the
fourth moment of a Nakagami-$m$ distributed RV, $E\left[ \left\vert
h\right\vert ^{4}\right] =\frac{m+1}{m}\Omega _{h}^{2}$.
\end{proof}

It is noted that \eqref{rhom} reduces to \eqref{Theo} for $m=1$. Expanding (\ref{rhom}) in a Taylor series for $\bar{\gamma}\rightarrow \infty $
and using (\ref{g1}), we obtain for $\alpha >-1$
\begin{equation}
g\left( \bar{\gamma}\right) =1-\rho \approx \frac{m}{\beta L}\bar{\gamma}%
^{-\left( \alpha +1\right) }+o\left( \bar{\gamma}^{-\left( \alpha +1\right)
}\right) .
\end{equation}
Next, we shed light onto the asymptotic behavior of relay selection with imperfect CSI
in Nakagami-$m$ fading.

\section{\label{DIV}Diversity Analysis}

\subsection{\label{OutHigh}High-SNR Analysis}

Here, we present a high SNR outage expression, which is used as a stepping
stone for deriving the diversity order of relay selection with imperfect CSI in
Nakagami-$m$ fading. For simplicity of exposition, this expression pertains
to the cases of outdated CSI and noisy CSI for Nakagami-$m$ fading, as well as FIR and IIR channel
prediction in Rayleigh fading; an expression for FIR and IIR channel
prediction in Nakagami-$m$ fading follows, likewise, from (\ref{FIRm}).

For sufficiently high values of $\bar{\gamma}$, we have%
\begin{equation}
\frac{\Gamma \left( m,\frac{m}{\bar{\gamma}}y\right) }{\Gamma \left(
m\right) }=\frac{\Gamma \left( m\right) -\mathcal{\gamma }\left( m,\frac{m}{%
\bar{\gamma}}y\right) }{\Gamma \left( m\right) }\approx \frac{\Gamma \left(
m\right) -\frac{\left( \frac{my}{\bar{\gamma}}\right) ^{m}}{m}}{\Gamma
\left( m\right) }=1-\frac{m^{m-1}y^{m}}{\Gamma \left( m\right) \bar{\gamma}%
^{m}}  \label{High_1b}
\end{equation}%
which holds based on the series representation of the lower incomplete Gamma
function, $\mathcal{\gamma }\left( s,x\right) $ \cite[Eq. (8.354/1)]%
{Gradshteyn}, in conjunction with the fact that as $x\rightarrow 0$, $%
\mathcal{\gamma }\left( s,x\right) \approx x^{s}/s$. Therefore, by
substituting (\ref{High_1b}) in (\ref{Final}) and setting $\rho =1-b\bar{%
\gamma}^{-a}$, as implied by (\ref{grd}), we obtain an alternative
expression for the outage probability as a function of the parameters $a$
and $b$ for high SNR as follows%
\begin{align}
P_{out}\approx & \left( \frac{m^{m-1}y^{m}}{\Gamma \left( m\right) \bar{%
\gamma}^{m}}\right) ^{N}+\sum_{j=0}^{\infty }\sum_{l=1}^{N}\frac{l\binom{N}{l%
}}{j!\Gamma \left( m\right) \Gamma \left( m+j\right) }\left( \frac{1-b\bar{%
\gamma}^{-a}}{b\bar{\gamma}^{-a}}\right) ^{n}\left( \frac{m^{m-1}T^{m}}{%
\Gamma \left( m\right) \bar{\gamma}^{m}}\right) ^{N-l}  \notag \\
& \times \left( 1-\frac{m^{m-1}T^{m}}{\Gamma \left( m\right) \bar{\gamma}^{m}%
}\right) ^{l}\left( \Gamma \left( m+j\right) -\Gamma \left( m+j,\frac{mT}{b%
\bar{\gamma}^{1-a}}\right) \right) G\left( m+j,a,b,l\right)  \label{High_3}
\end{align}%
where we have set%
\begin{equation}
G\left( m+j,a,b,l\right) =\sum_{i=0}^{l-1}\left( -1\right) ^{i}\binom{l-1}{i}%
\psi \left( m+j,1-b\bar{\gamma}^{-a},i\right) .  \label{G}
\end{equation}

\subsection{\label{DG}Diversity Order}

An important result derived from the high SNR analysis of Section \ref%
{OutHigh} is the diversity gain of the scheme under consideration, which is
summarized in the ensuing theorem.

\begin{theorem}
\label{DivCorr} The diversity gain of relay selection with imperfect CSI with
practical channel estimation and Nakagami-$m$ fading is given by%
\begin{equation}
G_{d}=\left\{
\begin{array}{c}
m\left[ a\left( N-1\right) +1\right] \text{, \ \ if }a<1 \\
\text{ \ \ \ \ \ \ \ \ \ \ }mN\text{, \ \ \ \ \ \ \ \ if }a\geq 1%
\end{array}%
\right.  \label{DivGain}
\end{equation}
\end{theorem}

\begin{proof}
The proof is given in Appendix \ref{DivAp}.
\end{proof}

\subsection{\label{DivPot}On The Diversity Potential of Relay Selection}

Based on the high SNR analysis of the previous section, interesting
results regarding the diversity order of relay selection can be obtained.
These results are presented below, for different types of channels and different
channel estimation techniques.

\subsubsection{Channel Estimation over Static Channels}

Let us first focus on the scenario where the channel estimates are obtained
via training in static channels. In this case, as can be seen from Table \ref%
{Table} the exponent $a$ can take any positive value, depending on how fast
the training power increases with SNR, with respect to the data transmission
power. In particular, it follows from Theorem \ref{DivCorr} that
full diversity is achieved by using a training power which increases with
SNR at least as fast as the data transmission power, i.e., $\alpha\geq0$. A slower increase with
SNR results in a decreased diversity order. Further details regarding the
latter argument are provided in Section \ref{NR}, via numerical examples.

\subsubsection{FIR Channel Prediction in Time-Varying Channels}

Here, we concentrate our attention on the case of FIR channel prediction;
note that the special case of outdated channel estimates is also included in
this scenario, by setting the number of predictor taps equal to one. As
shown in Table \ref{Table}, this case results in $a=0$. Interestingly, it
follows from \eqref{DivGain} that the diversity order equals the 
fading shape parameter, $m$, regardless of
the number of available relays. In other words, \textit{when the estimates
of time-varying channels are obtained via an FIR predictor, the diversity
order of relay selection reduces to that of the scheme where only a single
relay is available}. From another viewpoint, \textit{when relay selection is performed over
time-varying channels, its full diversity potential is completely lost,
unless a predictor with an infinitely large length is employed}. A study of
the latter case follows.

\subsubsection{IIR Channel Prediction in Time-Varying Channels}

As implied by Table \ref{Table} and Theorem \ref{DivCorr}, the diversity loss
of relay selection incurred by the time-varying nature of the underlying
channels can be recovered via IIR channel prediction. Nevertheless, given
that the full diversity order is recovered for $a\geq 1$, it follows that if
the power of the channel estimation pilot symbols grows with SNR as fast as
the data transmission power, i.e., $\alpha =0$, the resulting diversity
order is still lower than the maximum value. As a result, it is concluded
that \textit{in order to achieve full diversity in relay selection over
time-varying channels, IIR channel prediction is required, in conjunction
with a training power which increases faster than the data transmission
power, i.e.,} $\alpha >0$. The amount of training power required to achieve
full diversity is determined by the Doppler spread of the channel and the time difference between the consecutive noisy
channel observations.

\section{\label{NR}Numerical Results and Discussions}

\begin{figure}[tbp]
\centering
\includegraphics[keepaspectratio,width=%
\columnwidth]{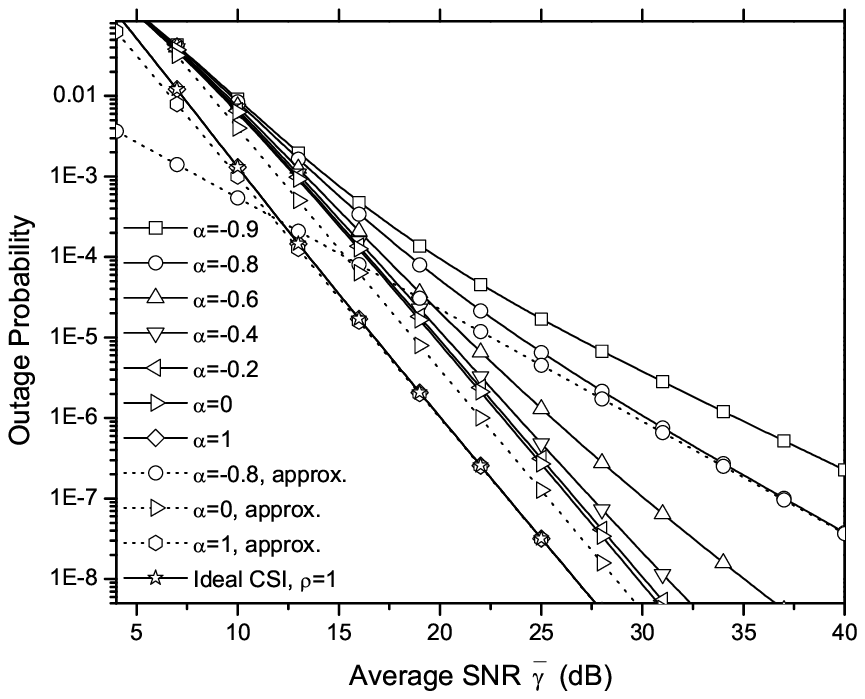}
\caption{Case of static channels: Outage probability of relay selection with
noise-impaired channel estimates in Rayleigh fading ($m=1$) versus
average SNR, assuming $N=3$ available relays, $T=1$, $\protect\beta =1$ and
several different values of $\protect\alpha $.}
\label{Staticm1}
\end{figure}
\begin{figure}[ptb]
\centering
\includegraphics[keepaspectratio,width=%
\columnwidth]{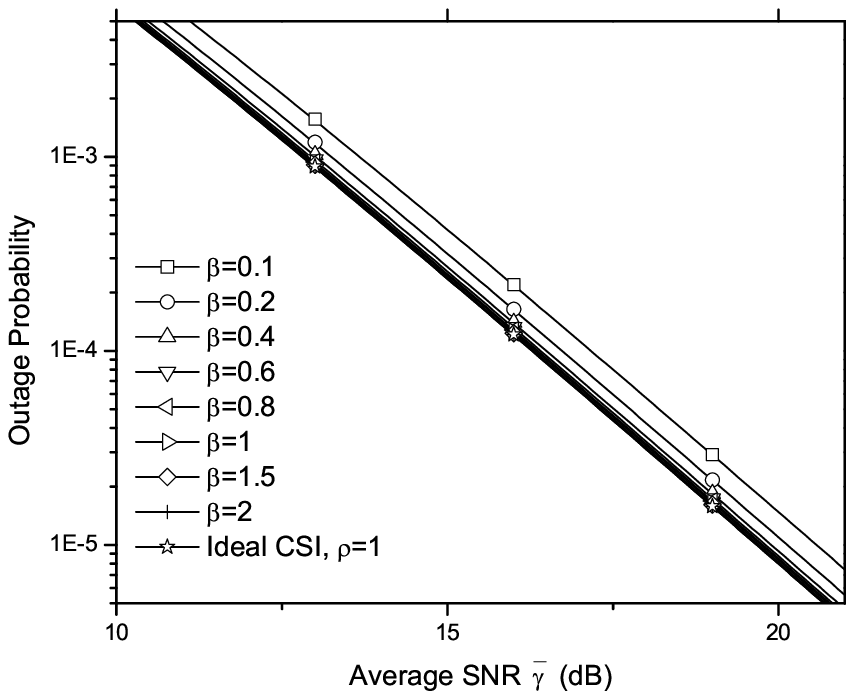}
\caption{Case of static channels: Outage probability of relay selection with
noise-impaired channel estimates in Rayleigh fading versus average SNR,
assuming $N=3$ available relays, $T=1$, $\protect\alpha=0$ and several
different values of $\protect\beta$.}
\label{Staticm2}
\end{figure}

Figs. \ref{Staticm1} and \ref{Staticm2} consider the case of static channels,
and illustrate the effect of noisy channel estimates on the outage
probability of relay selection over Nakagami-$m$ fading. Specifically, Fig. %
\ref{Staticm1} depicts results for the special case of Rayleigh fading, $N=3$
available relays and $\beta=1$, showing a significant dependence of the outage probability
on the parameter $\alpha$. Recall from \eqref{Ep} that
the parameters $\alpha$ and $\beta$ reflect the relation between the power
allocated to pilot symbols and the power used for data
transmission. It is observed that full diversity is achieved for any $%
\alpha\geq 0$, yet there exists a power gain loss compared to the perfect CSI case;
this power gain loss is recovered for higher values of $\alpha$, i.e., for $%
\alpha\geq 1$. In Fig. \ref{Staticm1}, we also observe the
accuracy of the high SNR approximations in \eqref{Cs13}, \eqref{Csa1},
and \eqref{Cs18} for $a<1$ ($\alpha<0$), $a\thickapprox 1$ ($%
\alpha\thickapprox 0$), and $a>1$ ($\alpha>0$), respectively.

The dependence of the outage probability on the ratio of the pilot power and data transmission power, $\beta$, is depicted in Fig. %
\ref{Staticm2}, for the case of static channels. Similarly as in Fig. \ref{Staticm1}, $N=3$ available relays
and Rayleigh fading ($m=1$) are assumed, while the $\alpha$ parameter is set
to $\alpha=0$. We notice a slight dependence of the outage probability on $%
\beta$ for a given $\alpha$, which becomes negligible as $\beta$ grows
large. Consequently, it is concluded that when the ratio of $%
\mathcal{E}_{p}$ and $\mathcal{E}_{d}$ is constant in the whole SNR
region, relay selection maintains its full diversity characteristics when
operating over noisy static channels; the corresponding loss in power gain
is noticeable only for small values of the ratio of $\mathcal{E}_{p}$ and $\mathcal{E}_{d}$.

\begin{figure}[tbp]
\centering
\includegraphics[keepaspectratio,width=\columnwidth]{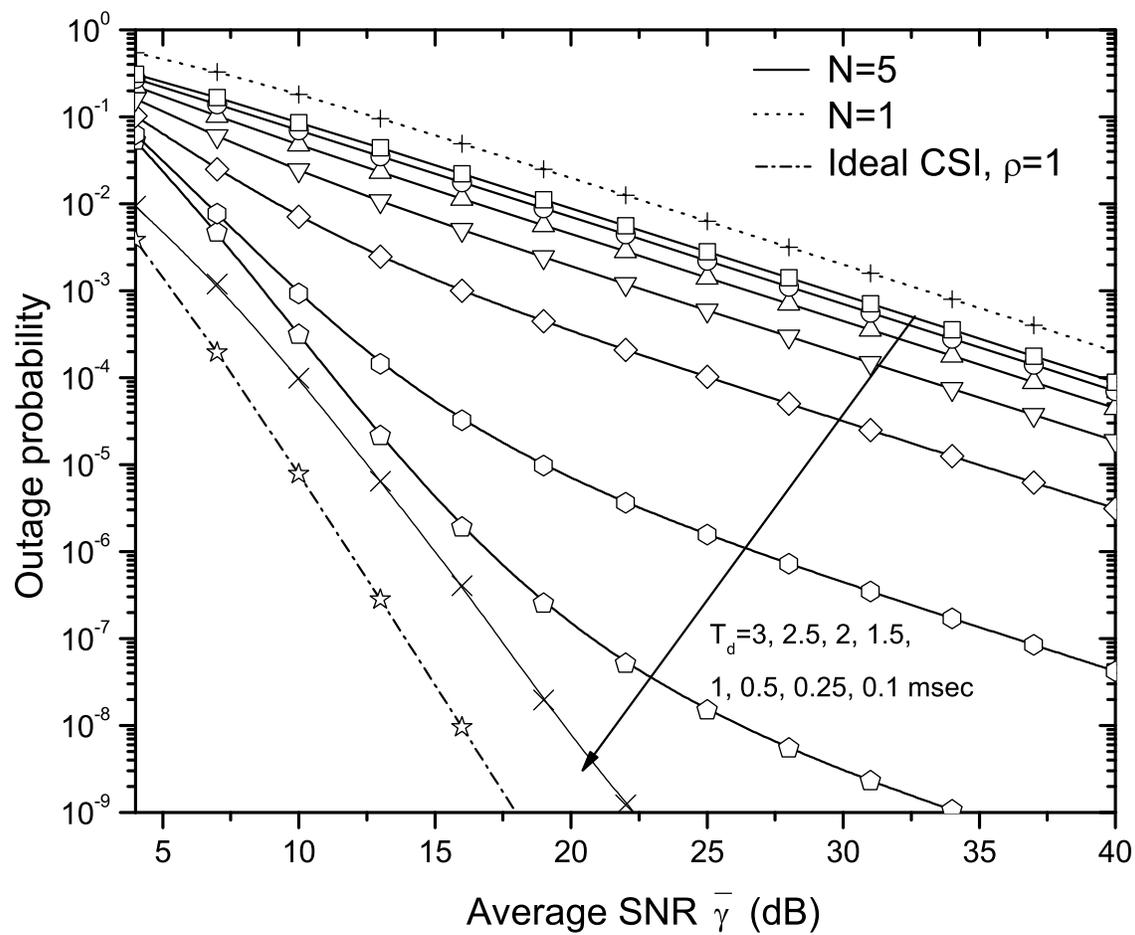}
\caption{Case of outdated CSI: Outage probability of relay selection in
Rayleigh fading versus average SNR, assuming $N=5$ relays, $%
T=1 $, maximum Doppler frequency $f_{d}=100$ Hz and several different values of $T_{d}$%
. }
\label{Outage1}
\end{figure}

\begin{figure}[ptb]
\centering
\includegraphics[keepaspectratio,width=\columnwidth]{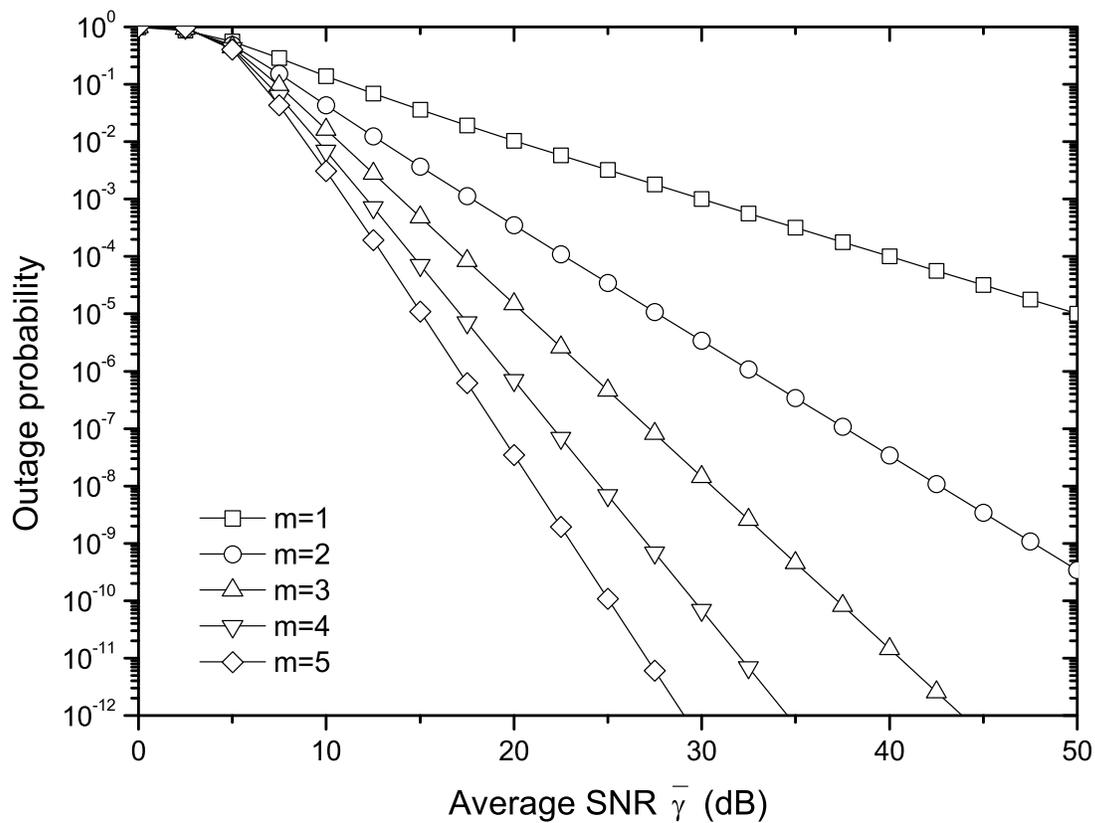}
\caption{Case of outdated CSI: Outage probability versus average SNR for
$\protect\rho=0.5$, $N=5$, $T=3$, and several values of the Nakagami-$m$
shape distribution parameter, $m$.}
\label{mdep}
\end{figure}

Figs. \ref{Outage1} and \ref{mdep} consider to the case of ideal but outdated
CSI, where the channel estimation is assumed noise-free yet it suffers from
feedback delay. Specifically, in Fig. \ref{Outage1} we assume the typical
scenario of a vehicle moving at $50$ km/h and receiving at a frequency of $%
2.4$ GHz, which
corresponds to a maximum Doppler frequency of approximately $f_{d}=100$ Hz.
Under this assumption, we illustrate the dependence of the corresponding
outage probability on the time interval between estimation updates, $T_{d}$,
for Rayleigh fading ($m=1$), $T=1$ and $N=5$.

We notice from Fig. \ref{Outage1} that the rate of estimation update
significantly affects the outage performance of relay selection, in the
sense that low update rates result in severe diversity and power gain losses.
This is in agreement with \eqref{DivGain} where, given that for outdated
channel estimates $a=0$ holds, the diversity order equals $m$ regardless of $%
T_{d}$. Nonetheless, it should be pointed out that for low values of $T_{d}$
the slope of the outage curves retains its full diversity characteristics in
the practical SNR range, and approaches $m$
only for infinitely high SNRs. This observation sheds light onto the
diversity potential of relay selection with outdated channel estimates since,
\textit{although it is impossible to achieve full diversity from a
theoretical perspective (i.e., when $\bar{\gamma}\rightarrow \infty $), it
is still possible to achieve full diversity in the practical SNR range, by
decreasing $T_{d}$}. Fig. \ref{Outage1} also demonstrates that for
relatively low channel estimation update rates (e.g., for $T_{d}=3$ msec),
relay selection cannot take advantage of the large number of available
relays, since the case of $N=5$ yields approximately the same
performance as that of no selection, i.e., $N=1$. Furthermore, it is worth mentioning that the outage probability
for the case where the mobile terminals are moving at the walking speed of $%
5$ km/h can be also extracted from Fig. \ref{Outage1}, by tenfolding the
corresponding values of $T_{d}$ (i.e., $T_{d}=30,25,...,1$ msec).

The outage probability dependence of relay selection with outdated CSI on
the Nakagami-$m$ parameter is depicted in Fig. \ref{mdep}. We assume five
participating relays ($N=5$) and an outage threshold SNR of $T=3$, while the
relation between $f_{d}$ and $T_{d}$ is set such that $\rho =0.5$. As
expected from \eqref{DivGain}, it is seen that increasing $m$ results in a
considerable outage probability decrease, accompanied by a shift of the
slope of the outage curves at high SNR. Note that in all cases the diversity
order equals $m$, as also corroborated by \eqref{DivGain}.

\begin{figure}[tbp]
\centering
\includegraphics[keepaspectratio,width=\columnwidth]{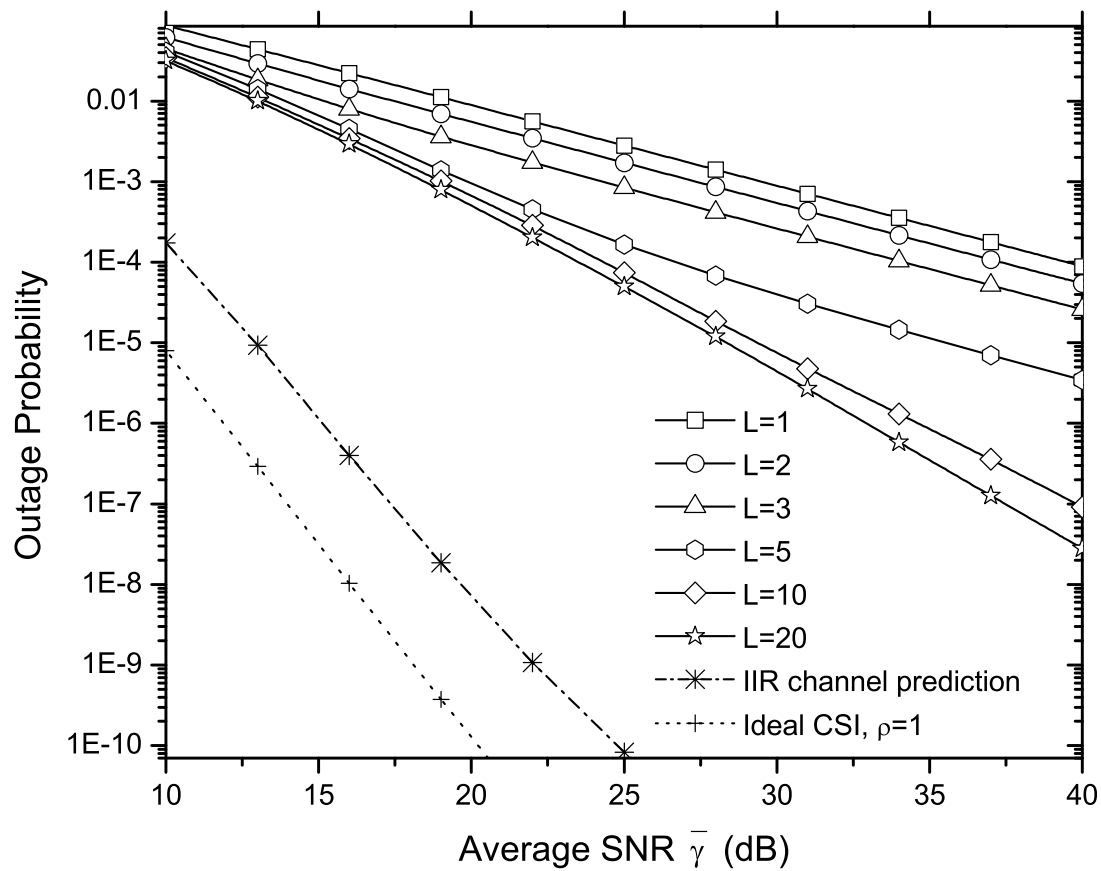}
\caption{Time-varying channels with FIR channel prediction: Outage
probability of relay selection in Rayleigh ($m=1$) fading versus average
SNR, assuming $N=5$ available relays, $T=1$, $T_{d}=3$ msec, $f_{d}=100$ Hz, $\protect\alpha %
=0$, $\protect\beta =1$, and several values of the channel predictor length,
$L$.}
\label{FIR}
\end{figure}

\begin{figure}[ptb]
\centering
\includegraphics[keepaspectratio,width=\columnwidth]{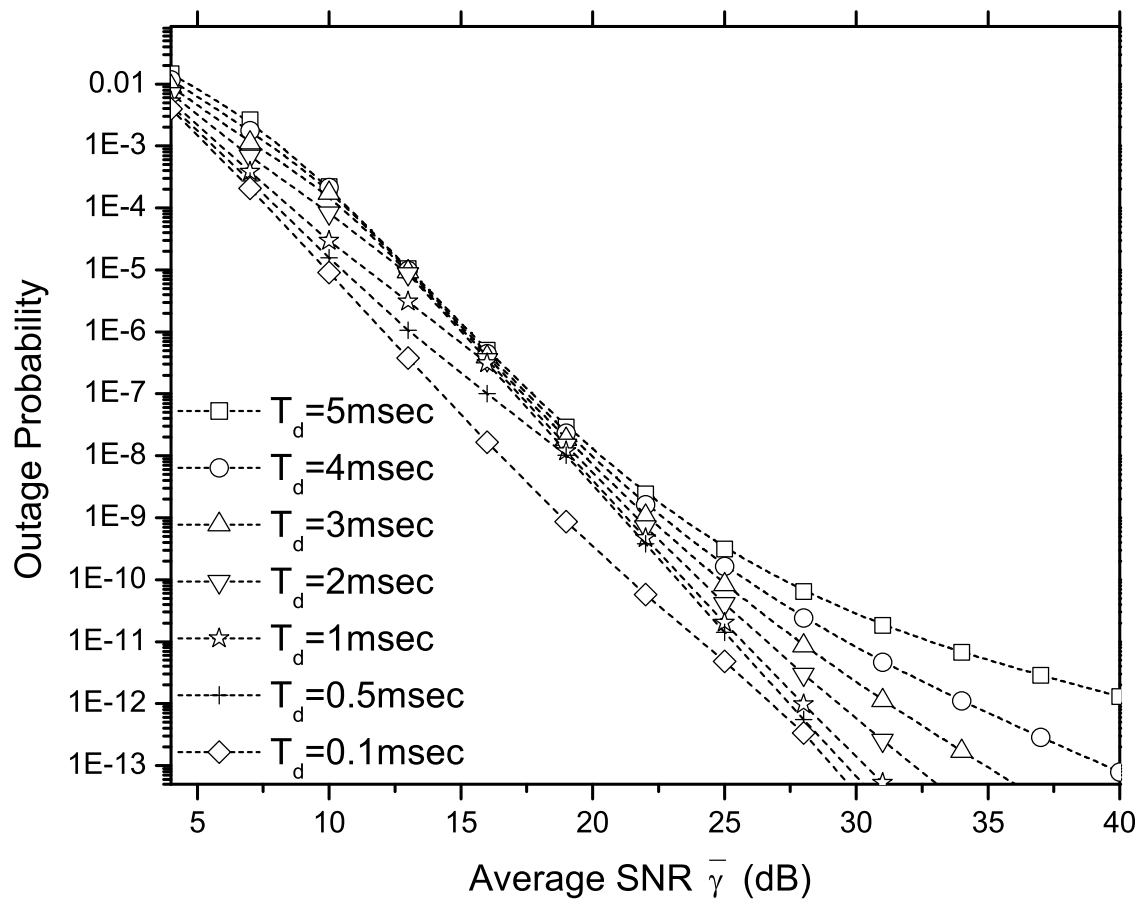}
\caption{Time-varying channels with IIR channel prediction: Outage
probability of relay selection in Rayleigh ($m=1$) fading versus average
SNR, assuming $N=5$ available relays, $T=1$, $\protect\alpha=0$, $\protect%
\beta=1$, and several values of $T_{d}$.}
\label{IIR}
\end{figure}

On the basis of the moving-vehicle scenario considered above, which corresponds to $f_{d}=100$ Hz, the case of
channel prediction in time-varying channels is treated in Figs. \ref{FIR}
and \ref{IIR}. In particular, Fig. \ref{FIR} depicts the outage probability
of relay selection in Rayleigh fading ($m=1$) for several values of
the channel predictor length, $L$, including the case of $L\rightarrow
\infty $ which corresponds to IIR channel prediction and serves here as
benchmark. As demonstrated in Fig. \ref{FIR}, by increasing the number of
predictor coefficients in FIR channel prediction the outage probability experiences a
power gain increase, yet no diversity gain increase is seen for low values
of $L$. On the contrary, an increase in the diversity gain is attained
through IIR channel prediction, as shown in the ensuing, Fig. \ref{IIR}.

Fig. \ref{IIR} illustrates the outage probability of relay selection for
IIR channel prediction, when operating over Rayleigh fading. We notice a
high dependence of the outage probability on the value of $T_{d}$, which corresponds to the
time difference among the consecutive time instances in the infinite-length
channel predictor. In particular, we notice that full diversity is achieved
for small values of $T_{d}$, while for larger values of $T_{d}$ the
diversity characteristics of relay selection are lost, as expected from \eqref{DivGain}
and Table \ref{Table}.

Finally, the achievable diversity order versus $a$ for the unified imperfect
CSI model, where the cases of noisy channel estimation and CSI imperfection
due to time-varying channels are incorporated, is plotted in Fig. \ref%
{FigDivOrd}. As expected, we notice a linear increase of the diversity order
for $0 \leq a \leq 1$ and constant diversity order for $a \geq 1$, which
equals $mN$.

\begin{figure}[tbp]
\centering
\includegraphics[keepaspectratio,width=\columnwidth]{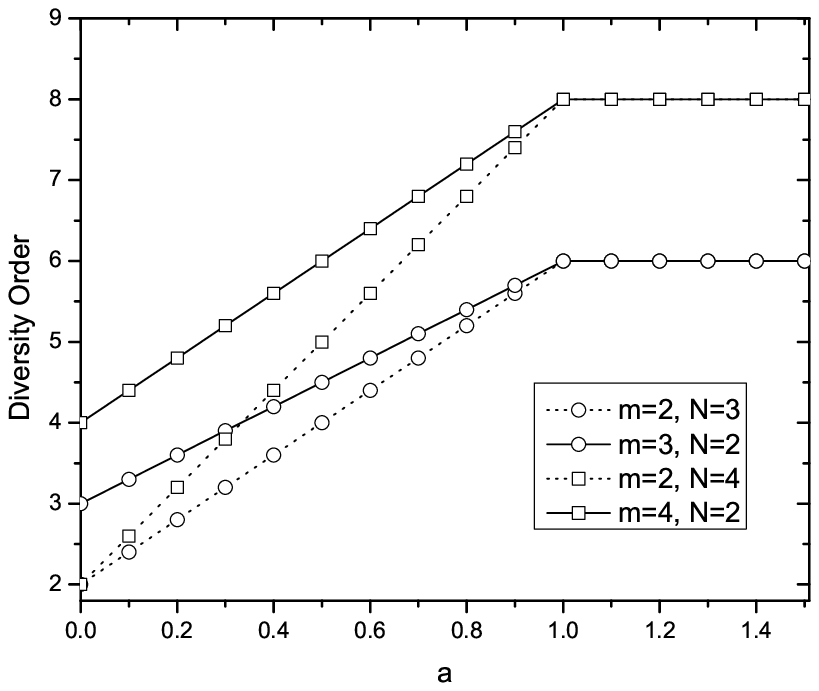}
\caption{Diversity order of relay selection with imperfect CSI in Nakagami-$%
m $ fading as a function of the exponent parameter $a$, for several
combinations of the number of available relays, $N$, and the fading shape
parameter, $m$. Applies to all the considered imperfect CSI scenarios.}
\label{FigDivOrd}
\end{figure}

In fact, Fig. \ref{FigDivOrd} sheds an interesting light onto our general assessment
regarding the diversity order of relay selection in Nakagami-$m$ fading,
which is as follows. The Nakagami-$m$ fading model assumes multiple
scatterers in each link, causing an \textquotedblleft \textit{internal
diversity}\textquotedblright ~phenomenon of order $m$, which is independent
of the channel estimation quality. On the other hand, the presence of
multiple available relays offers the potential for additional,
\textquotedblleft \textit{external diversity}\textquotedblright\, yet this additional
diversity strongly depends on the quality of channel estimation, as
reflected by $a$. Consequently, we notice from Fig. \ref{FigDivOrd} that for any non-prime
diversity order of $mN$, the value of $m$ is more important than $N$ for the overall diversity order for any $\alpha <1$; if $a\geq
1 $, $m$ and $N$ affect the diversity order in exactly the same way.

\section{Conclusions}

We presented an assessment of the diversity potential of relay selection with practical channel estimation techniques, in Nakagami-$m$ fading. The considered channel estimation techniques include the cases of estimation in noisy static channels; estimation in noiseless time-varying channels, and estimation in noisy time-varying channel with the aid of FIR and IIR channel prediction. A closed-form expression for the outage probability of relay selection with imperfect CSI was provided, as a function of the correlation coefficient, $\rho$, between the actual and the estimated channel values. Capitalizing on this outage expression, our principal inference was that the diversity order of relay selection is determined by the relative speed of convergence of $\rho$ to one, compared to the speed that the SNR converges to infinity.

\appendices

\section{\label{ApIIR}IIR Channel Prediction}


Let us consider the process $y_{l}=h_{l}+n_{l}
$,
where $h_{l}$ and $n_{l}$ denote the channel value and the corresponding
noise component at time instance $l$, respectively, with $l=1,...,\infty $.
The variance of the prediction error is derived as \cite{pred2}
\begin{equation}
\sigma _{e}^{2}=\exp \left( T_{d}\int_{-1/\left(2T_{d}\right)}^{1/\left(2T_{d}\right)}\ln \left[
\mathcal{S}_{hh}\left( e^{j2\pi fT_{d}}\right) +\frac{N_{0}}{\mathcal{E}_{p}}%
\right] df\right) -\frac{N_{0}}{\mathcal{E}_{p}}  \label{Appb1}
\end{equation}%
where $N_{0}/\mathcal{E}_{p}$ is the variance of the noise component and $%
\mathcal{S}_{hh}\left( f\right) $ denotes the Fourier transform of $\rho
_{h}\left( \tau \right) $. Therefore, considering that the spectrum of $%
\mathcal{S}_{hh}\left( f\right) $ is band-limited by $f_{d}$, (\ref{Appb1})
yields%
\begin{equation}
\sigma _{e}^{2}=\exp \left( T_{d}\int_{-f_{d}}^{f_{d}}\ln \left[ \mathcal{S}%
_{hh}\left( e^{j2\pi fT_{d}}\right) +\frac{N_{0}}{\mathcal{E}_{p}}\right]
df\right) \left( \frac{N_{0}}{\mathcal{E}_{p}}\right) ^{1-2f_{d}T_{d}}-\frac{%
N_{0}}{\mathcal{E}_{p}}.  \label{Appb3}
\end{equation}%
Assuming $f_{d}>0$, (\ref{rhosqIIR}) is obtained from (\ref{Appb3}), (\ref%
{rho1}), and (\ref{Ep}).

\section{\label{ApI}Derivation of the Auxiliary Function $I\left( \cdot
,\cdot ,\cdot ,\cdot ,\cdot \right) $}

Using the alternative representation of the incomplete Gamma function shown
in \cite[Eq. (8.352/4)]{Gradshteyn}, and applying the multinomial theorem,
we obtain%
\begin{align}
\Gamma ^{j}\left( m,\beta x\right) & =\left( \left( m-1\right) !\right)
^{j}\exp \left( -j\beta x\right) \left( \sum_{i=0}^{m-1}\frac{\beta ^{i}x^{i}%
}{i!}\right) ^{j}  \notag \\
& =\Gamma ^{j}\left( m\right) \exp \left( -j\beta x\right) \sum_{\substack{ %
n_{0},n_{1},...,n_{m-1}=0  \\ n_{0}+n_{1}+...+n_{m-1}=j}}^{j}\Gamma \left(
j+1\right) \prod\limits_{i=0}^{m-1}\frac{\left( \frac{\beta ^{i}}{i!}%
x^{i}\right) ^{n_{i}}}{\Gamma \left( n_{i}+1\right) }.  \label{Gamma2}
\end{align}%
Therefore, (\ref{I1}) yields%
\begin{align}
I\left( \mu ,\alpha ,m,\beta ,j\right) & =\sum_{\substack{ %
n_{0},n_{1},...,n_{m-1}=0  \\ n_{0}+n_{1}+...+n_{m-1}=j}}^{j}\Gamma
^{j}\left( m\right) \Gamma \left( j+1\right) \left( \prod\limits_{i=0}^{m-1}%
\frac{\left( \frac{\beta ^{i}}{i!}\right) ^{n_{i}}}{\Gamma \left(
n_{i}+1\right) }\right) \int_{0}^{\infty }x^{\mu
+\sum\limits_{i=0}^{m-1}in_{i}}
e^{-\left( a+j\beta \right) x}
dx  \notag \\
& =\frac{\Gamma ^{j}\left( m\right) \Gamma \left( j+1\right) }{\left(
a+j\beta \right) ^{\mu +1}}\sum_{\substack{ n_{0},n_{1},...,n_{m-1}=0  \\ %
n_{0}+n_{1}+...+n_{m-1}=j}}^{j}\left( \prod\limits_{i=0}^{m-1}\frac{\left(
\frac{\beta ^{i}}{i!\left( a+j\beta \right) ^{i}}\right) ^{n_{i}}}{\Gamma
\left( n_{i}+1\right) }\right) \left( \mu
+\sum\limits_{i=0}^{m-1}in_{i}\right) !.  \label{I2}
\end{align}

\section{\label{DivAp}Diversity Analysis in Nakagami-$m$ fading}

The following Lemma provides a high-SNR investigation of function $%
G\left( m+j,a,b,l\right) $, allowing for a simplification of (\ref{High_3}).

\begin{lemma}
\label{Gg}For sufficiently
high $\bar{\gamma}$, the function $G\left(
m+j,a,b,l\right) $, defined in (\ref{G}), decays proportionally to $\bar{%
\gamma}^{-\left( alm+aj\right) }$, i.e.,%
\begin{equation}
\lim_{\bar{\gamma}\rightarrow \infty }G\left( m+j,a,b,l\right) ~\sim \bar{%
\gamma}^{-\left( alm+aj\right) }.
\end{equation}

\begin{proof}
It follows from (\ref{coef}) that the function $\psi \left( m+j,1-b\bar{%
\gamma}^{-a},i\right) $ takes the following form%
\begin{equation}
\psi \left( m+j,1-b\bar{\gamma}^{-a},i\right) =\left( \frac{\bar{\gamma}^{a}%
}{b}+i\right) ^{-j}\psi \left( m,1-b\bar{\gamma}^{-a},i\right) .
\label{psi_recip}
\end{equation}%
Therefore, for deriving the decay exponent of $G\left( m+j,a,b,l\right) $ it
suffices to evaluate the decay exponent of $G\left( m,a,b,l\right) $. After
substituting (\ref{coef}) into (\ref{G}) and algebraic manipulations, the function $G\left(
m,a,b,l\right) $ can be expressed as%
\begin{equation}
G\left( m,a,b,l\right) =\frac{\mu _{0}\bar{\gamma}^{-am}\left( 1+\mu _{1}%
\bar{\gamma}^{a}+\mu _{2}\bar{\gamma}^{2a}+...+\mu _{\left(
\sum_{i=1}^{l-1}i\right) \left( m-1\right) }\bar{\gamma}^{\left(
\sum_{i=1}^{l-1}i\right) \left( m-1\right) a}\right) }{\left( 1+\bar{\gamma}%
^{a}\right) ^{2m-1}\left( 2+\bar{\gamma}^{a}\right) ^{3m-2}\times ...\times
\left( l-1+\bar{\gamma}^{a}\right) ^{lm-\left( l-1\right) }}  \label{D2}
\end{equation}%
where $\mu _{\zeta }$, $\zeta \in \left\{ 0,...,\frac{l}{2}\left( l-1\right)
\left( m-1\right) \right\} $, are constants. It is observed 
from (\ref%
{D2}) that the dominant term of $G\left( m,a,b,l\right) $ in the high SNR
regime decays in proportion to $\bar{\gamma}^{-d_{G}}$, where $d_{G}$ is
given by%
\begin{align}
d_{G}& =am-\left( \sum_{i=1}^{l-1}i\right) \left( m-1\right)
a+a\sum_{i=2}^{l}\left[ im-\left( i-1\right) \right] = alm. \label{D3}
\end{align}%
The proof then follows from (\ref{D3}) and (\ref{psi_recip}), by applying
the binomial expansion to the first term of the right hand side of (\ref%
{psi_recip}).
\end{proof}
\end{lemma}

For the derivation of the diversity order, we simplify \eqref{High_3} for different values of $a$, as shown below.

\subsubsection{Case of $a<1$}

In this case, the second argument of $\Gamma \left( m+n,mT/\left( b\bar{%
\gamma}^{1-a}\right) \right) $ in (\ref{High_3}) tends to zero as $\bar{\gamma}\rightarrow
\infty $. Hence, simplifying the incomplete Gamma function similarly as in (%
\ref{High_1b}), we obtain%
\begin{equation}
\Gamma \left( m+j\right) -\Gamma \left( m+j,\frac{mT}{b\bar{\gamma}^{1-a}}%
\right) \approx \frac{m^{m+j}y^{m+j}}{\left( m+j\right) b^{m+j}\bar{\gamma}%
^{\left( 1-a\right) \left( m+j\right) }}.  \label{Gammam}
\end{equation}%
Therefore, from (\ref{High_3}) and (\ref{Gammam}) we obtain an expression
for the outage probability in the form of%
\begin{equation}
P_{out}\left( y\right) \approx \left( \frac{m^{m-1}T^{m}}{\Gamma \left( m\right) \bar{\gamma%
}^{m}}\right) ^{N}+\mathcal{C}  \label{Cs11}
\end{equation}%
where%
\begin{equation}
\mathcal{C}=\sum_{j=0}^{\infty }\sum_{l=1}^{N}\frac{l\binom{N}{l}\left(
\frac{1-b\bar{\gamma}^{-a}}{b\bar{\gamma}^{-a}}\right) ^{n}\left( \frac{%
m^{m-1}T^{m}}{\Gamma \left( m\right) \bar{\gamma}^{m}}\right) ^{N-l}}{%
j!\Gamma \left( m\right) \Gamma \left( m+j\right) }\frac{\left( 1-\frac{%
m^{m-1}y^{m}}{\Gamma \left( m\right) \bar{\gamma}^{m}}\right)
^{l}m^{m+j}T^{m+j}}{\left( m+j\right) b^{m+j}\bar{\gamma}^{\left( 1-a\right)
\left( m+j\right) }}G\left( m+j,a,b,l\right) .  \label{Cs_12}
\end{equation}%
The following lemma investigates the high SNR behavior of $\mathcal{C}$.

\begin{lemma}
\label{Gc2}For sufficiently high $\bar{\gamma}$ and $a<1$, the quantity $%
\mathcal{C}$ defined in (\ref{Cs_12}) is approximated by reducing the
sums to single terms corresponding to $j=0$ and $l=N$, respectively, and
decays in proportion to $\bar{\gamma}^{-m\left[ a\left( N-1\right) +1\right]
}$, i.e.,%
\begin{equation}
\lim_{\bar{\gamma}\rightarrow \infty }\mathcal{C}~\sim \bar{\gamma}^{-m%
\left[ a\left( N-1\right) +1\right] }.  \label{LemGc2}
\end{equation}
\end{lemma}

\begin{proof}
It follows from \textit{Lemma} \ref{Gg} that in the high SNR region the
quantity inside the summation of (\ref{Cs_12}) is analogous to $\bar{\gamma}%
^{d_{\mathcal{C}}}$, where $d_{\mathcal{C}}$ is given by%
\begin{equation}
d_{\mathcal{C}}=an-m\left( N-l\right) -\left( 1-a\right) \left( m+n\right)
-\left( alm+an\right) =\left( 1-a\right) \left( ml-m-n\right) -mN.
\label{dc}
\end{equation}%
Since $a<1$, (\ref{dc}) is maximized over the set of non-negative integers $%
j $ for $j=0$, regardless of $l$; this value of $j$ corresponds thus to the
dominant term in the outer sum of (\ref{Cs_12}). Consequently, (\ref{dc})
reduces to%
\begin{equation}
d_{\mathcal{C}}=m\left[ \left( 1-a\right) \left( l-1\right) -N\right] .
\label{dc2}
\end{equation}%
Since $a<1$ and $1\leq l\leq N$, it follows from (\ref{dc2}) that the
dominant term in the inner sum of (\ref{Cs_12}) corresponds to $l=N$. Hence,
setting $l=N$ to (\ref{dc2}) completes the proof.
\end{proof}

Therefore, considering the fact that the first term in \eqref{Cs11} 
decays in proportion
to $\bar{\gamma}^{-mN}$, it follows from \textit{Lemma }\ref{Gc2} that for $%
a<1$ and $N>1$, the dominant term in (\ref{Cs11}) is $\mathcal{C}$,
where only the term with $j=0$ and $l=N$ is relevant, so that (\ref{Cs11}) reduces to%
\begin{equation}
P_{out}\approx N\frac{m^{m-1}T^{m}}{\Gamma ^{2}\left( m\right) b^{m}\bar{%
\gamma}^{\left( 1-a\right) m}}G\left( m,a,b,N\right) .  \label{Cs13}
\end{equation}%
Eq. (\ref{Cs13}) represents a high SNR expression for the outage probability
for $a<1$.

\subsubsection{Case of $a>1$}

Let us now assume that $a$ is larger than one, and does not lie in the
proximity of one. The case where $a$ approaches unity will be considered
separately in Section \ref{a1}. For $a>1$, we have%
\begin{equation}
\lim_{\bar{\gamma}\rightarrow \infty }\left[ \Gamma \left( m+j\right)
-\Gamma \left( m+j,\frac{mT}{b\bar{\gamma}^{1-a}}\right) \right] =\Gamma
\left( m+j\right) .  \label{Gammam2}
\end{equation}%
Hence, following the same procedure as for proving \textit{Lemma }\ref{Gc2},
it is concluded that the $l$th-order term within the double summation in (%
\ref{High_3}) decays proportionally to $\bar{\gamma}^{-m\left[ N+l\left(
a-1\right) \right] }$, irrespective of $j$. Consequently, since $a>1$ the
negative decay exponent of the second term of (\ref{High_3}) is higher than $%
mN$ for any $l\geq 1$. This implies that the dominant term in (\ref{High_3})
is the first term for high SNRs, yielding%
\begin{equation}
P_{out}\approx \left( \frac{m^{m-1}}{\Gamma \left( m\right) }\right)
^{N}\left( \frac{T}{\bar{\gamma}}\right) ^{mN}.  \label{Cs18}
\end{equation}%
Eq. (\ref{Cs18}) represents the asymptotic outage expression in high SNR for
$a>1$.

\subsubsection{\label{a1}Case of $a\thickapprox 1$}

The scenario where $a$ lies in the proximity of one is treated separately,
since in this case the second argument of $\Gamma \left( m+j,mT/\left( b\bar{%
\gamma}^{1-a}\right) \right) $ converges very slowly (to either zero or
infinity) as $\bar{\gamma}\rightarrow \infty $, hence the approximations in (%
\ref{Gammam}) and (\ref{Gammam2}) do not hold for practical SNR values. As a
result, the transition from the $a<1$ case to the $a>1$ case experiences a
discontinuity in terms of the (practical) high SNR approximation of the
outage probability, as $a$ approaches unity. This discontinuity is bridged
through the outage expression presented below. Recall from Section \ref%
{Static} that the case of $a=1$ corresponds to the common scenario of
channel estimation in noisy static channels, where the power allocated to
pilot symbols equals the power allocated to data transmission.

Since $a\thickapprox 1$, let us assume that 
$\bar{\gamma}%
^{1-a}$ approaches a non-zero finite constant as $\bar{\gamma}\rightarrow
\infty $, i.e.,%
\begin{equation}
\lim_{\substack{ a\rightarrow 1  \\ \bar{\gamma}\rightarrow \infty }}\bar{%
\gamma}^{1-a}=\lambda ,\text{ \ }0<\lambda <\infty .  \label{lambda}
\end{equation}%
This allows us to evaluate the integral $I\left( m+j-1,m/\left( b\bar{\gamma}%
^{1-a}\right) ,m,m/\bar{\gamma},i\right) $, shown in (\ref{I1}), as follows.

\begin{itemize}
\item Let us assume $0<x<\infty $
. Then, using \cite[Eq. (8.352/7)]{Gradshteyn} we have for high SNR%
\begin{equation}
\Gamma ^{i}\left( m,\frac{m}{\bar{\gamma}}x\right) \approx \Gamma ^{i}\left(
m\right) .  \label{Gamma}
\end{equation}

\item Let $x\rightarrow \infty $. In this case, (\ref{Gamma}) does not hold.
However, it follows from L'Hospital's rule, as well as from the fact that $%
b\lambda $ is finite, that%
\begin{equation}
\lim_{x\rightarrow \infty }\left[ x^{m+j-1}\exp \left( -\frac{m}{b\lambda }%
x\right) \Gamma ^{i}\left( m,\frac{m}{\bar{\gamma}}x\right) \right] =0.
\label{Gamma02}
\end{equation}
\end{itemize}

Therefore, it follows from (\ref{Gamma}) and (\ref{Gamma02}) that for high
SNRs%
\begin{equation}
I\left( m+j-1,\frac{m}{b\bar{\gamma}^{1-a}},m,\frac{m}{\bar{\gamma}}%
,i\right) \approx \int_{0}^{\infty }x^{m+j-1}\exp \left( -\frac{mx}{b\lambda
}\right) \Gamma ^{i}\left( m\right) dx=\Gamma ^{i}\left( m\right) \Gamma
\left( m+j\right) \left( \frac{b\lambda }{m}\right) ^{m+j}.  \label{Cs12}
\end{equation}

By combining (\ref{High_3}), (\ref{coef}), (\ref{I2}), (\ref%
{lambda}), (\ref{Cs12}), \cite[Eq. (0.15.4)]{Gradshteyn}, \cite[Eq. (8.310/1)]{Gradshteyn}, \cite[Eq. (8.350/2)]{Gradshteyn}, and the infinite series representation of the exponential function, we arrive after some manipulations at
\begin{equation}
P_{out}\approx \left( N+1\right) \left( \frac{m^{m-1}}{\Gamma \left(
m\right) }\right) ^{N}\left( \frac{T}{\bar{\gamma}}\right) ^{mN}.
\label{Csa1}
\end{equation}%
Eq. (\ref{Csa1}) represents a high SNR approximation of the outage
probability for the case where $a$ lies in the neighborhood of one. \emph{Theorem} \ref{DivCorr} follows then directly from \eqref{Cs13}, \eqref{Csa1}, and \eqref{Cs18}.

\newpage
\begin{table}[tbp]
\caption{Parameters $a$ and $b$ for the considered imperfect CSI scenarios.}
\label{Table}\centering%
\begin{tabular}{|l|l|l|}
\hline
\emph{Case} & $a$ & $b$ \\ \hline
Outdated CSI & $0$ & $1-\rho $ \\ \hline
Noisy CSI & $\alpha +1$ & $m/\left( \beta L\right) $ \\ \hline
FIR Channel Prediction & $0$ & $1-\text{\b{u}}_{h}^{H}\left. \text{\b{R}}%
^{-1}\right\vert _{N_{0}=0}\text{\b{u}}_{h}$ \\ \hline
IIR Channel Prediction & $\left( \alpha +1\right) \left(
1-2f_{d}T_{d}\right) $ & $\exp \left( T_{d}\int_{-f_{d}}^{f_{d}}\ln \left[
S_{hh}\left( e^{j2\pi fT_{d}}\right) \right] df\right) \beta ^{-\left(
1-2f_{d}T_{d}\right) }$ \\ \hline
\end{tabular}%
\end{table}
\newpage

\newpage

\newpage

\newpage

\newpage

\newpage

\newpage

\begin{thebibliography}{99}
\bibitem{dohler} M. Dohler and Y. Li, \textquotedblleft \emph{Cooperative
communications: Hardware, channel \& PHY,}\textquotedblright\ Wiley \& Sons,
2010.

\bibitem{uysal} M. Uysal (Ed.), \textquotedblleft \emph{Cooperative
communications for improved wireless network transmission: Frameworks for
virtual antenna array applications\textquotedblright }. IGI-Global, 2009.

\bibitem{uysal2} M. M. Fareed and M. Uysal, \textquotedblleft On relay
selection for decode-and-forward relaying\textquotedblright, \emph{IEEE
Trans. Wireless Commun.}, vol. 8, no. 7, p. 3341-3346, July 2009.

\bibitem{beres} E. Beres and R. Adve, \textquotedblleft Selection
cooperation in multi-source cooperative networks,\textquotedblright\ \emph{%
IEEE Trans. Wireless Commun.}, vol. 7, pp. 118-127, Jan 2008.

\bibitem{zhao} Y. Zhao, R.S. Adve and T.J. Lim, \textquotedblleft Improving
amplify-and-forward relay networks: optimal power allocation versus
selection\textquotedblright, \emph{IEEE Trans. Wireless Commun.}, vol. 6,
no. 8., pp. 3114-3123, Aug. 2007.

\bibitem{mehta} V. Shah, N. B. Mehta, and R. Yim, \textquotedblleft Relay
selection and data transmission throughput tradeoff in cooperative
systems\textquotedblright , \emph{IEEE Global Telecommunications Conference
(Globecom)}, Honolulu, USA, Dec. 2009.

\bibitem{mehta2} B. Medepally, and N. B. Mehta, \textquotedblleft Voluntary
energy harvesting relays and selection in cooperative wireless
networks,\textquotedblright\ \emph{IEEE Trans. on Wireless Commun.}, vol.9,
pp.3543-3553, Nov 2010.

\bibitem{vicario} J. L. Vicario, A. Bel, J. A. Lopez-Salcedo, and G. Seco,
\textquotedblleft Opportunistic relay selection with outdated CSI: Outage
probability and diversity analysis,\textquotedblright\ \emph{IEEE Trans.
Wireless Commun.}, vol. 8, pp. 2872-2876, June 2009.

\bibitem{outCSI} D. S. Michalopoulos, H. A. Suraweera, G. K. Karagiannidis,
and R. Schober, \textquotedblleft Relay selection with outdated channel
estimates,\textquotedblright\ \emph{IEEE Global Communications Conference
(Globecom) 2010 }Miami, FL, USA.

\bibitem{outCSINak} D. S. Michalopoulos, N. D. Chatzidiamantis, R. Schober
and G. K. Karagiannidis, \textquotedblleft Relay selection with outdated
channel estimates in Nakagami-$m$ fading\textquotedblright , to be presented
at \emph{IEEE International Conference on Communications (ICC)}, 2011.

\bibitem{outCSIJ} D. S. Michalopoulos, H. A. Suraweera, G. K. Karagiannidis,
and R. Schober, \textquotedblleft Amplify-and-forward relay selection with
outdated channel estimates\textquotedblright , submitted to \emph{IEEE
Trans. on Commun.}

\bibitem{torabi} M. Torabi and D. Haccoun, \textquotedblleft Capacity
analysis of opportunistic relaying in cooperative systems with outdated
channel information\textquotedblright , \emph{IEEE Commun. Letters}, vol.
14, pp. 1137-1139, Dec 2010.

\bibitem{soysa} H. A. Suraweera, M. Soysa, C. Tellambura, and H. K. Garg,
\textquotedblleft Performance analysis of partial relay selection with
feedback delay\textquotedblright , \emph{IEEE Signal Processing Letters},
vol.17, pp.531-534, Jun 2010.

\bibitem{gayan} G. Amarasuriya, C. Tellambura, and M. Ardakani,
\textquotedblleft Feedback delay effect on dual-hop MIMO AF relaying with
antenna selection,\textquotedblright \emph{IEEE Global Telecommunications
Conference (GLOBECOM)}, 2010.

\bibitem{gayan2} G. Amarasuriya, M. Ardakani, and C. Tellambura,
\textquotedblleft Output-threshold multiple-relay-selection scheme for
cooperative wireless networks,\textquotedblright~ \emph{IEEE Trans. Veh.
Technol.}, vol.59, pp.3091-3097, Jul 2010.

\bibitem{seyfi} M. Seyfi, S. Muhaidat and J. Liang, \textquotedblleft
Performance analysis of relay selection with feedback delay and channel
estimation errors,\textquotedblright\ \emph{IEEE Signal Proc. Letters}, vol.
18, Jan 2011.

\bibitem{jafar} M. J. Taghiyar, S. Muhaidat and J. Liang, \textquotedblleft
On the performance of pilot symbol assisted modulation for cooperative
systems with imperfect channel estimation,\textquotedblright \emph{IEEE
Wireless Communications and Networking Conference (WCNC)}, 2010.

\bibitem{cherlin} G. Cherlin, \emph{Model Theoretic Algebra Selected Topics}%
. Lecture notes in mathematics 521. Springer-Verlag, 1976.

\bibitem{Jakes} W. C. Jakes, \textquotedblleft \emph{Microwave mobile
communication\textquotedblright }, J. Wiley\&Sons, NY, 1974.

\bibitem{Gradshteyn} I. S. Gradshteyn and I. M. Ryzhik, \textit{Table of
integrals, series, and products,} New York, Academic Press, 7th edition,
2007.

\bibitem{makhoul} J. Makhoul, \textquotedblleft Linear prediction: A
tutorial overview\textquotedblright, \emph{IEEE Proceedings, }Vol. 63, No.4,
Apr 1975.

\bibitem{dowton} F. Downton, \textquotedblleft Bivariate exponential
distributions in reliability theory\textquotedblright . J\emph{ournal of the
Royal Statistics Society}, Series B, vol 32, No. 3 (1970), pp. 408-417.

\bibitem{nakagami} M. Nakagami, \textquotedblleft The -distribution-A
general formula of intensity distribution of rapid
fading,\textquotedblright\ in \textit{Statistical Methods in Radio Wave
Propagation}, W. C. Hoffman, Ed. Oxford, U.K.: Pergamon, 1960, pp. 3--36

\bibitem{Bivariate_Nakagami} J. Reig, L. Rubio and N. Cardona,
\textquotedblleft Bivariate Nakagami-$m$ with arbitrary fading
parameters,\textquotedblright\ \textit{Electron. Lett.}, vol. 38, no. 25,
Dec. 2002.

\bibitem{pred2} B. Picinbono and J.-M. Kerilis, \textquotedblleft Some
properties of prediction and interpolation errors\textquotedblright, \emph{%
IEEE Trans. on Acoustics, Speech and Signal Processing, } vol.36,
pp.525-531, Apr 1988.







\end{thebibliography}
\end{document}